\documentclass[traditabstract]{aa}  
\usepackage{graphicx}
\usepackage{txfonts}
\graphicspath{{figures/}}

\usepackage{hyperref}
\hypersetup{breaklinks=true,colorlinks=true,linkcolor=blue,citecolor=blue,urlcolor=blue}

\usepackage{natbib,twoopt}
\bibpunct{(}{)}{;}{a}{}{,}             %% natbib format for A&A and ApJ

\defcitealias{dr3}{DR3}

\newcommand{\hlr}{HLR} %{R$_{50}$}
\newcommand{\hmr}{HMR} %{R$^{M}_{50}$}
\newcommand{\mstar}{M$_{\star}$}
\newcommand{\msun}{M$_{\odot}$}
\newcommand{\nmustar}{stellar mass surface density}
\newcommand{\mustar}{$\Sigma_{\star}$} %{$\mu_{\star}$}
\newcommand{\mustargal}{$\Sigma^{galaxy}_{\star}$} %{$\mu_{\star}$}
\newcommand{\mustarhlr}{$\Sigma^{HLR}_{\star}$} %{$\mu_{\star}$}
\newcommand{\mustarunits}{M$_{\odot}$ pc$^{-2}$}
\newcommand{\ngal}{661}
\newcommand{\tg}[1]{t$_{#1}$}
\newcommand{\tgin}[1]{t$_{#1}^{In}$}
\newcommand{\tgout}[1]{t$_{#1}^{Out}$}
\newcommand{\mmbin}[4]{(#1 $-$ #2, #3 $-$ #4) $\lbrack$log \mstar, log \mustar$\rbrack$}
\newcommand{\mbin}[2]{(#1 - #2) [log \mstar]}
\newcommand{\mdbin}[2]{(#1 - #2) [log \mustar]}
\newcommand{\pow}[1]{10$^{#1}$}

\newcommand{\tage}{$age$}
\newcommand{\agem}{$\langle$log \tage$\rangle_{M}$}
\newcommand{\agel}{$\langle$log \tage$\rangle_{L}$}

\newcommand{\lmstar}{log M$_{*}$ [M$_{\odot}$]} 
\newcommand{\lmustar}{log \mustar [\mustarunits]}
\newcommand{\gagem}{$\bigtriangledown \langle \log\,\, age \rangle_{M}$}

\newcommand{\pycasso}{{\sc p}y{\sc casso}}
\newcommand{\stl}{{\scshape starlight}}

\newcommand{\sfig}{v20.q057.d22a500.ps03.k1.mE.CCM.Bgstf6e}
\newcommand{\sprop}{Mini}

\begin{document}

\titlerunning{Cosmic time scales}
%\authorrunning{Garc\'ia-Benito et al.}

\title{The spatially resolved star formation history of CALIFA galaxies:} 
\subtitle{Cosmic time scales}

\author{R.~Garc\'ia-Benito\inst{\ref{inst1}}
        \and R.~M.~Gonz\'alez Delgado\inst{\ref{inst1}}
        \and E.~P\'erez\inst{\ref{inst1}}
        \and R.~Cid Fernandes\inst{\ref{inst2}}
        \and C.~Cortijo-Ferrero\inst{\ref{inst1}}
        \and R.~L\'opez Fern\'andez\inst{\ref{inst1}}
        \and A.~L.~de~Amorim\inst{\ref{inst2}}
        \and E.~A.~D~Lacerda\inst{\ref{inst2}}
        \and N.~Vale~Asari\inst{\ref{inst2}}
        \and S.~F.~S\'anchez\inst{\ref{inst3}}
          \and CALIFA\inst{4}
}

\institute{Instituto de Astrof\'isica de Andaluc\'ia (CSIC), P.O. Box 3004, 18080 Granada, Spain\label{inst1} 
(\email{rgb@iaa.es})
   \and Departamento de F\'isica, Universidade Federal de Santa Catarina, P.O. Box 476, 88040-900, Florian\'opolis, SC, Brazil\label{inst2}
   \and Instituto de Astronom\'ia, Universidad Nacional Auton\'oma de M\'exico, A.P. 70-264, 04510 M\'exico D.F., M\'exico\label{inst3}
   \and \href{http://califa.caha.es}{http://califa.caha.es}\label{inst4}
}

\date{}

\abstract{
This paper presents the mass assembly time scales of nearby galaxies observed by CALIFA at the 
3.5m telescope in Calar Alto. We apply the fossil record method of the stellar populations 
to the complete sample of the 3rd CALIFA data release, with a total of \ngal\ galaxies, covering 
stellar masses from \pow{8.4} to \pow{12} \msun\ and a wide range of Hubble types. 
We apply spectral synthesis techniques to the datacubes and process 
the results to produce the mass growth time scales and mass weighted ages, 
from which we obtain temporal and spatially resolved information in seven bins of galaxy 
morphology (E, S0, Sa, Sb, Sc, and Sd) and six bins of stellar mass and \nmustar. 
We use three different tracers of the spatially resolved star formation history (mass assembly 
curves, ratio of half mass to half light radii, and mass-weighted age gradients) to test if  
galaxies grow inside-out, and its dependence with galaxy stellar mass, 
\nmustar, and morphology. Our main results are as follows: 
{\em (a)} The innermost regions of galaxies assemble their mass at an earlier time than regions 
located in the outer parts; this happens at any given stellar mass (\mstar), \nmustar\ (\mustar), 
or Hubble type, including the lowest mass systems in our sample.
{\em (b)} Galaxies present a significant diversity in their characteristic formation epochs
for lower-mass systems. This diversity shows a strong dependence of the mass assembly 
time scales on \mustar\ and Hubble type in the lower-mass range (\pow{8.4} to \pow{10.4}), 
but a very mild dependence in higher-mass bins.
{\em (c)} The lowest half mass radius (\hmr) to half light radius (\hlr) ratio is found for galaxies 
between \pow{10.4} and \pow{11.1} \msun, where galaxies are 25\% smaller in mass than in light. 
Low-mass galaxies show the largest ratio with \hmr/\hlr $\sim$ 0.89. Sb and Sbc galaxies present 
the lowest \hmr/\hlr\ ratio (0.74). The ratio \hmr/\hlr\ is always, on average, below 1, indicating 
that galaxies grow faster in mass than in light.
{\em (d)} All galaxies show negative \agem\ gradients in the inner 1 \hlr. 
The profile flattens (slope less negative) with increasing values of \mustar. There is no 
significant dependence on \mstar\ within a particular \mustar\ bin, except for the lowest 
bin, where the gradients becomes steeper.
{\em (e)} Downsizing is spatially preserved as a function of \mstar\ and \mustar, but it is 
broken for E and SO where the outer parts are assembled in later epochs than Sa galaxies.
These results suggest that independently of their stellar mass, \nmustar,\ and morphology, 
galaxies form inside-out on average.
}

% \abstract{}{}{}{}{} 
% 5 {} token are mandatory
 
%\abstract


\maketitle
%
%-------------------------------------------------------------------

%%%%%%%%%%%%%%%%%%%%%%%%%%%%%%%%%%%%
%%%%%%%%%%%%% Section %%%%%%%%%%%%%%
%%%%%%%%%%%%%%%%%%%%%%%%%%%%%%%%%%%%
\section{Introduction}

Tracing how galaxies grow their stellar mass as a function of cosmic time is key to understanding how galaxies 
form and evolve. Nowadays, a two-phase scenario is proposed for the formation of massive galaxies in which  
ellipticals (E) form first a dense core at high redshift, and subsequently grow their outer envelope 
\citep{vanDokkum:2010, Patel:2013}. This mass growth occurs with a significant increase of their size 
\citep{Trujillo:2006, Buitrago:2008} that is produced mainly by dry mergers \citep{Naab:2009, Oser:2010, Hilz:2013}. 
As the central core of E, spiral (S) galaxies form also their bulges at high redshift mainly by mergers. 
Mergers cannot play a significant role in the formation of disks since major mergers can destroy disks 
\citep{Toomre:1972, Hopkins:2009}, although recent simulations show cases where new disks soon regrow 
\citep{Baugh:1996,Steinmetz:2002,Springel:2005,Athanassoula:2016}.
Thin disks are thought to be formed from haloes with more quiescent merging history 
that accrete cool gas and gradually form stars with a radial distribution around the central bulge set by the 
angular momentum distribution of the halo \citep{Neistein:2006, Birnboim:2007, Dutton:2010}. Both pictures 
suggest an inside-out view for the formation of a galaxy \citep{Mo:1998, AvilaReese:2000, Pilkington:2012a, Aumer:2013}. 

The overall process for the formation of galaxies is more complex, because the mass growth is not driven solely by 
gas supply. Other mechanisms like feedback from supernovae or Active Galactic Nucleus (AGN) are 
required to decouple the star formation from 
the gas accretion and to stop the growth of the galaxy \citep{Croton:2006, Dekel:2009, Lilly:2013}. Furthermore, the 
morphological transformation following the formation of a spheroidal component can stabilize the gas disk against 
fragmentation and stop the growth of the galaxy \citep{Martig:2009, Genzel:2014, GonzalezDelgado:2015}.

In contrast, many spirals like the Milky Way (MW) do not harbor a classical bulge \citep{Fisher:2010, diMatteo:2014}, and 
instead they have a central thick disk \citep{Haywood:2013, Haywood:2015} that may have formed by clumpy instabilities in 
the star forming disk \citep{Elmegreen:2008}. Some samples of galaxies at a redshift of $\sim$2  show clumpy star forming disks 
\citep{Elmegreen:2007, Genzel:2008, ForsterScheiber:2011, Wuyts:2013}, which suggest the formation of an early build up of 
pseudo-bulges by secular evolution. These results provide evidence for a modification of the inside-out formation scenario. 
In this line, \citet{vanDokkum:2013} analyzed a sample of MW-like spirals at $z \sim$2.5 and found that the mass growth took 
place in a fairly uniform radial way, since the size - stellar mass (\mstar) relation at $z \sim$2.5 is similar to the 
relation at $z = 0$. Moreover, other mechanisms, such as stellar migration, bar-induced gas inflows, and angular momentum loss 
due to reorientation of the disk, can alter the inside-out view of the formation of a galaxy \citep{Aumer:2014}.

Observational evidence in favor of the inside-out growth of star forming galaxies at high redshift already exists. 
For example, \citet{Nelson:2016a, Nelson:2012} have found that the H$\alpha$ distribution of star forming galaxies at 
$z \sim 1$ is more extended than the stellar continuum, consistent with an inside-out assembly of the galactic disks. 
Other evidence comes from the radial mass structure of galaxies at 0.5 $< z <$ 2.5 that shows that the half mass radius 
(\hmr) is on average 25\% smaller than the half light radius (\hlr). This result shows that galaxies are more 
compact in mass than in light, as a consequence of the inside-out mass growth of galaxies. The larger differences 
between the two radii occur in the most massive disk galaxies \citep{Szomoru:2010, Szomoru:2012}, in favor again of an 
inside-out formation scenario for disk galaxies.

In the local universe, further evidence of the inside-out formation scenario for spirals comes from the UV extended 
disks \citep{GildePaz:2005,  GildePaz:2007} detected by GALEX \citep{Martin:2005}, and the radial distribution of the 
specific star formation rate \citep{MunozMateos:2007}. However, recently it has been shown that spirals grow in size at 
a rate which is 0.35 times the rate at which they grow in mass \citep{Pezzulli:2015}.

Theoretically, an inside-out formation process is expected for disks where they should increase in size while they 
grow in mass. Since the stars are distributed in a rotating disk and the specific angular momentum increases outwards, 
the inner parts should form earlier than the outer parts \citep{Larson:1976, Brook:2006, Brook:2012, Pilkington:2012a}. 

This topic has achieved significant progress in the last years thanks to the ability of integral field spectroscopy 
(IFS) surveys (e.g., CALIFA \citep{califa}, MaNGA \citep{Bundy:2015}) and the fossil record method \citep{Tinsley:1968}
to recover the information of the spatially resolved star formation history of nearby galaxies, and to derive the mass 
assembly history as a function of radial distance. CALIFA is well suited for these studies thanks to the large 
field of view and the spatial resolution of the instrument. The radial structure of the stellar ages and mass-growth curves 
can be obtained out to distances larger than 2 \hlr\ and with a spatial sampling of $\sim$0.1-0.2 \hlr. 
Moreover, the large sample of galaxies of the CALIFA survey that covers all Hubble types \citep{Walcher:2014}, and 
galaxy stellar mass from 10$^9$ to $\sim$10$^{12}$ \msun\ \citep{GonzalezDelgado:2015} allow us to test the inside-out 
scenario for E, S0, and spirals from Sa to Sd, and low- to high-mass galaxies.

From the CALIFA survey, we have already obtained interesting results that favor the inside-out formation scenario. 
They are as follows: 1) From the mass assembly history of $\sim$ 100 galaxies \citep{Perez:2013}, we found that galaxies with 
$M_\star >$ 10$^{10}$ \msun\ grow inside-out, and the signal of downsizing is spatially preserved, with both 
inner and outer regions growing faster for massive galaxies. 2) From the ratio of the \hmr\ to the \hlr\ we found 
that these galaxies are more compact in mass than in light, supporting again the inside-out scenario 
\citep{GonzalezDelgado:2014a, GonzalezDelgado:2015}. 3) From the radial profiles of the stellar ages of $\sim$ 300 
galaxies, we found negative age gradients for early- and late-type galaxies, confirming again the inside-out 
formation for E and spiral galaxies \citep{GonzalezDelgado:2015}. This conclusion was also sustained by 
\citet{SanchezBlazquez:2014} for a similar study based in a sub-sample of face-on disk CALIFA galaxies. 
4) Using a completely different methodology, the study of the gas abundance gradients also suggests this 
inside-out scenario \citep{Sanchez:2012, Sanchez:2014, SanchezMenguiano:2016, Marino:2016}.

Nonetheless, recent results from the MaNGA survey find more divergence in their conclusions: 1) From the 
mass assembly history, \citet{IbarraMedel:2016} obtained that the innermost regions formed stars on average earlier 
than the outermost ones, supporting an inside-out formation scenario for $\sim$ 500 galaxies with \mstar\  
from 10$^{9.3}$ to 10$^{11.2}$ \msun. Their conclusion is independent of galaxy color, specific star 
formation rate (sSFR), and morphological type. 2) By using a similar set of data, \citet{Zheng:2017} found 
slight negative age gradients, that point to an inside-out formation process for $\sim$ 1100 galaxies, 
half of which have Sersic index $n <$ 2.5 (disk-like), and the other half of the sample have $n >$ 2.5 
(elliptical-like). 3) In contrast, for a similar set of data, but using different methods 
to retrieve the stellar population properties \citet{Goddard:2016} found a positive age gradient for the 
early-type galaxies, suggesting an outside-in formation process for these galaxies. They both did full 
spectral fitting, but with different codes; \citet{Zheng:2017} with \stl\ \citep{CidFernandes:2005}, 
and \citet{Goddard:2016} with {\scshape firefly} \citep{Wilkinson:2015}. They also use different 
stellar population models. Thus, their conclusions differ and seem to depend on the method of the analysis 
and the tracer (mass growth curve vs. age gradients) used to test the formation process.

In this paper, we analyze the complete set from the 3rd CALIFA data release \citep{dr3}, and test the results 
from three different tracers: 1) Mass growth curves; 2) ratio of \hmr\ to \hlr; 3) age radial profiles  
(weighted in mass) of the stellar populations. This analysis will allow us to complement our previous results 
for a large sample of galaxies. Furthermore, in previous studies 
\citep{Perez:2013, GonzalezDelgado:2014a, GonzalezDelgado:2015}, we did not conclusively prove that galaxies 
with M$_\star < 10^{10} M_\star$ form inside-out due to the poor statistics in this range of stellar mass. 
Now, the CALIFA extension sample will test the inside-out scenario for a large sample of galaxies, using a 
systematic analysis in the context of the three different diagnostic tracers mentioned above, as a function 
of Hubble type, stellar mass (\mstar), and \nmustar\ (\mustar).

This paper is organized as follows. In Sect. \ref{Sec:data}, we describe the observations and data. In Sect. 
\ref{Sec:method} we summarize the stellar population analysis, methodology, and sample distribution. Section 
\ref{Sec:mgh} presents the two-dimensional (2D) radial stellar mass growth maps, the spatial averaged mass assembly and 
characteristic mass assembly epoch results. Section \ref{Sec:hmr_hlr} discusses the ratio of half-mass to 
half-light radii as function of Hubble type, \mstar, and \mustar. Section \ref{Sec:mage} presents the 
mass-weighted age radial profiles. In Sect. \ref{Sec:discussion} we discuss the global mass assembly time scales, 
their multivariate dependence and the mass-weighted age profiles both as a function of \hlr\ and \hmr. Finally, 
Sect. \ref{Sec:conclusions} summarizes the results of the paper and our main conclusions. Throughout we assume a 
flat $\Lambda$CDM cosmology with $\Omega_{M}$ = 0.3, $\Omega_{\Lambda}$ = 0.7, 
and H$_{0}$ = 70 km s$^{-1}$ Mpc$^{-1}$. 

%--------------------------------------------------------------------
\section{Data and sample\label{Sec:data}}

\subsection{Data}
CALIFA \citep[the  Calar  Alto  Legacy  Integral  Field  Area  survey,][]{califa} is an Integral Field Spectroscopy (IFS) 
survey designed to obtain spatially resolved spectra for~600 galaxies in the local Universe. Observations were carried out 
with the 3.5  m telescope at Calar Alto observatory with the Postdam Multi Aperture Spectrograph \citep[PMAS,][]{Roth:2005} in 
the PPaK mode \citep{Verheijen:2004}. PPak contains a bundle of 331 science fibers, each of 2.7\arcsec in diameter, and a 
71\arcsec $\times$ 64\arcsec field of view \citep[FoV;][]{Kelz:2006}. A three-pointing dithering is used for each object 
in order to reach a filling factor of 100\% across the entire FoV. The observations were planed to observe each galaxy 
with two different overlapping setups. The low-resolution setup (V500; R$\sim$ 850) covers from 3745 to 7500 \AA\ with 
a spectral resolution of $\sim$ 6 \AA\ FWHM, while the medium resolution setup (V1200; R $\sim$ 1650) covers the blue 
part in the range 3400-4840 \AA\ with spectral resolution of $\sim$ 2.3 \AA\ FWHM. The spatial sampling of the datacubes 
is 1\arcsec/spaxel with an effective spatial resolution of $\sim$ 2.6\arcsec FWHM.

\subsection{Sample}
The CALIFA mother sample was drawn from the SDSS DR7 \citep{Abazajian:2009} photometric galaxy catalog to obtain a 
representative and statistically significant sample of galaxies in the nominal redshift range 0.005 $<$ $z$ $<$ 0.03, 
fully described in \citet{Walcher:2014}. 

We have used the full sample of the final CALIFA DR3 \citep[][hereafter DR3]{dr3}, with a total of 646 galaxies observed 
in the V500 grating. The DR3 release is the combination of two samples. The CALIFA Main Sample (MS) consists of 
galaxies belonging to the CALIFA mother sample, with a total of 542 in the V500 included in the DR3.
The rest of the galaxies belong to the CALIFA Extension Sample, a set of galaxies observed within the CALIFA collaboration 
as part of different ancillary science projects (see \citetalias{dr3}). They were reduced with the v2.2 CALIFA pipeline. 
The total number of datacubes in the V500 is 700, after adding all the galaxies from the ancillary science projects 
that have SDSS-DR7 images. We excluded type-1 Seyferts and galaxies that show merger or 
interaction features. This leaves a final sample of \ngal\ galaxies.

DR3 galaxies were morphologically classified by five members of the collaboration through visual inspection of the SDSS
$r$-band images, averaging the results as described in \citet{Walcher:2014}. We performed a similar classification procedure 
for the objects from the ancillary science projects.

As in previous works, we group the galaxies in seven morphology bins: E (120 galaxies), S0 (78, including S0 and S0a), 
Sa (94, including Sa and Sab), Sb (100), Sbc (93), Sc (110, including Sc and Scd), and Sd (66, including Sd, Sm, and Irr).

%%%%%%%%%%%%%%%%%%%%%%%%%%%%%%%%%%%%
%%%%%%%%%%%%% Section %%%%%%%%%%%%%%
%%%%%%%%%%%%%%%%%%%%%%%%%%%%%%%%%%%%
\section{Stellar population analysis\label{Sec:method}}

%%% Subsection %%%
\subsection{Method of analysis}

%----------------------------------------------------------------- 
\begin{figure}
\centering
%\resizebox{0.31\textwidth}{!}{\includegraphics{masshub_hist_\sfig}}
%\resizebox{0.31\textwidth}{!}{\includegraphics{massmcorsd_hist_\sfig}}
%\resizebox{0.31\textwidth}{!}{\includegraphics{mcorsdhub_hist_\sfig}}
\resizebox{\hsize}{!}{\includegraphics{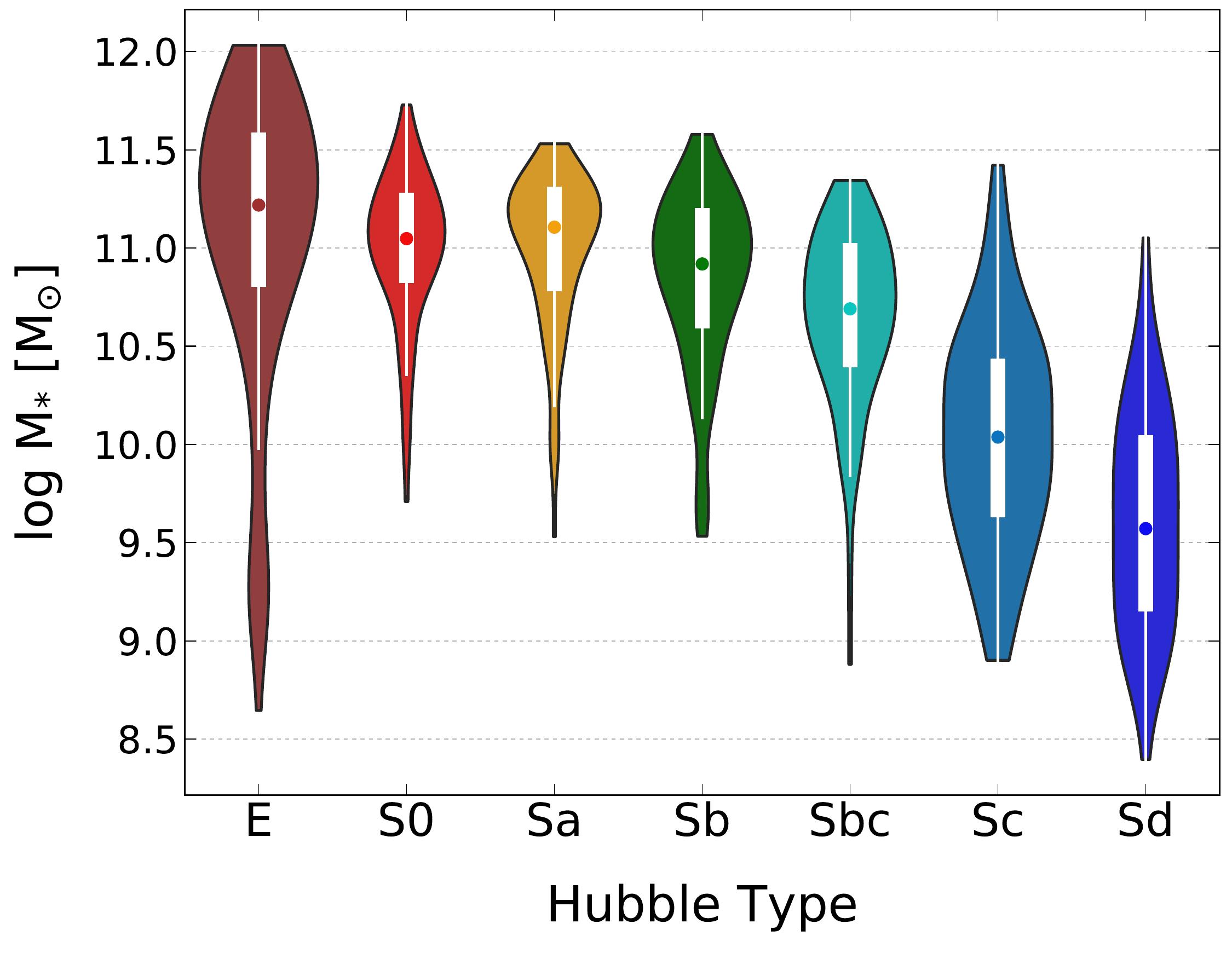}} 
\resizebox{\hsize}{!}{\includegraphics{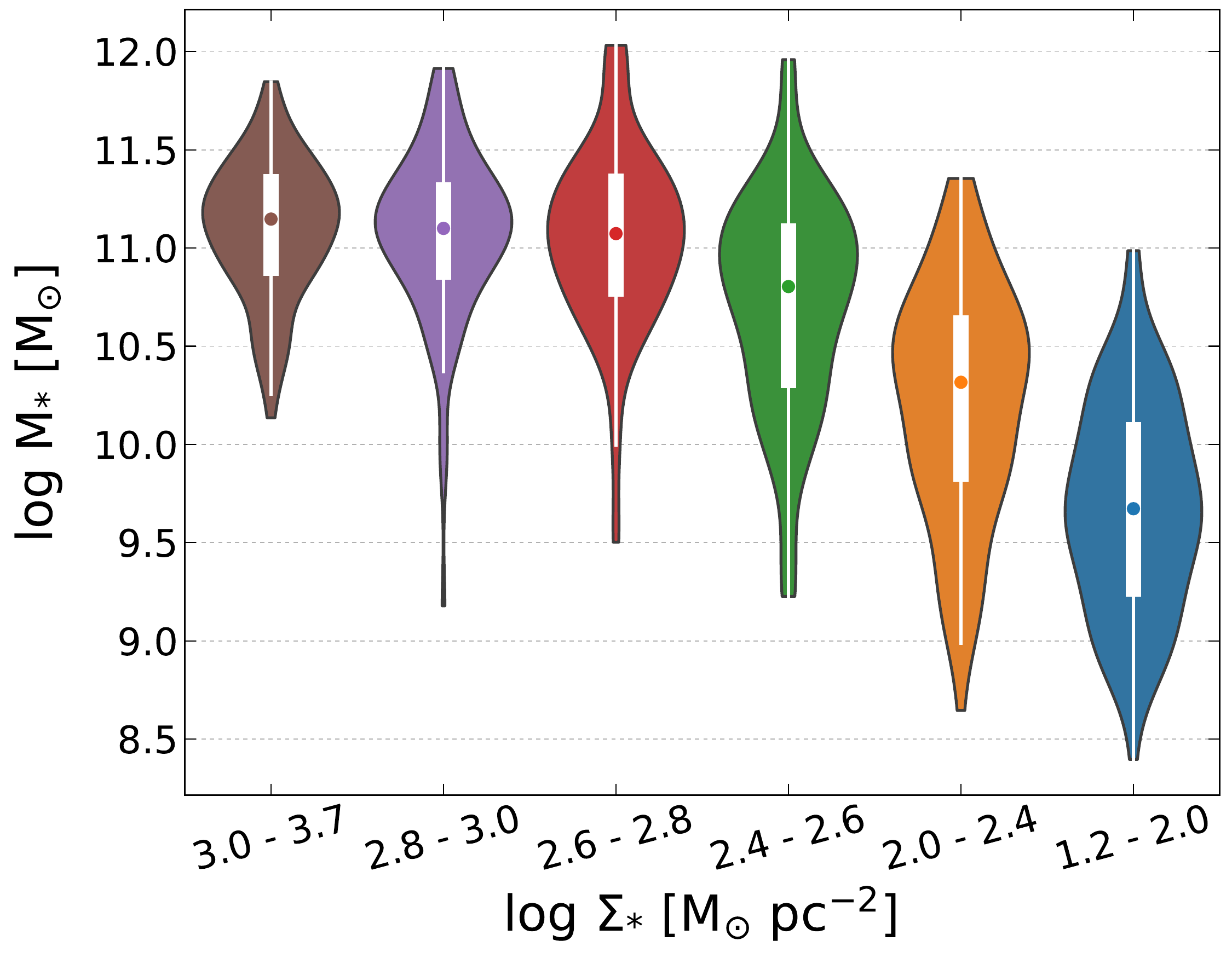}} 
\resizebox{\hsize}{!}{\includegraphics{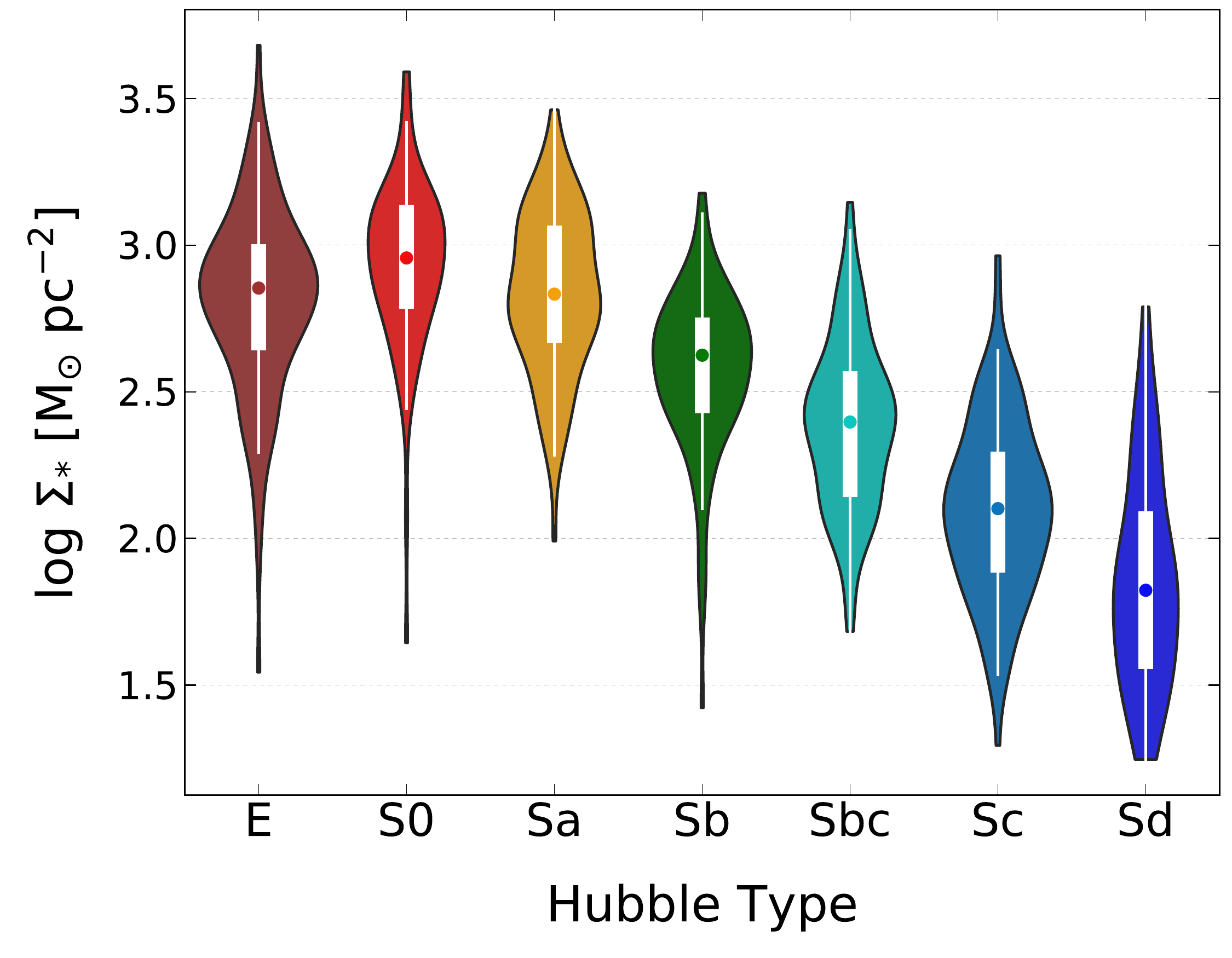}}
\caption{Violin plots of the stellar masses for each Hubble type (upper panel), stellar masses for 
each \nmustar\ bin (middle panel), and \nmustar\ for each Hubble type (bottom panel). Its 
corresponding box plot showing the interquartile range is plotted inside each violin plot. The 
inner dot in the box plot represents the median of the distribution. The values of the \nmustar\ 
were obtained at 1 \hlr\ (see Sect. \ref{Sec:mustar}). 
The width of the "violins" are scaled by the number of observations in each bin.}
\label{Fig:massmcorsdhub}
\end{figure}
%-----------------------------------------------------------------

The stellar population properties are derived from the datacubes following the same methodology 
applied in previous works \citep{Perez:2013, CidFernandes:2013, CidFernandes:2014, GonzalezDelgado:2014a, 
GonzalezDelgado:2014b, GonzalezDelgado:2015, GonzalezDelgado:2016}. In short, the work-flow comprises three 
major steps. First, the preprocessing is done via {\scshape qbick}, which includes spatial masking for 
foreground and background sources and masking of bad wavelength elements (large and small errors, 
bad pixels, ...). All spaxels within the isophote level where the average signal-to-noise ratio 
$\geq$ 3 (S/N) are spatially binned (if required) via Voronoi binning\footnote{Measured in a 90 
\AA\ window centered at 5635\AA\ rest-frame.} \citep{Cappellari:2003} so the final spectra 
achieve S/N $\gtrsim$ 20.
The coadded error spectra are corrected taking into account the correlation given by 
\citet{dr3}\footnote{See appendix of \citet{dr2} for a description of the procedure.}. 
In the second step, the resulting spectra are fitted with \stl\ using the cluster 
at the Instituto de Astrof\'isica de Andaluc\'ia. Finally, the output is packed and processed 
using \pycasso\footnote{The Python CALIFA \stl\ Synthesis Organizer.} 
\citep[de Amorim et al. 2017, in preparation]{CidFernandes:2013}.

The main difference with previous works is the updated base used in \stl\ to perform the full 
spectral fit. It consists of a combination of 254 SSP. For ages younger than 60 Myr we 
use the GRANADA models of \citet{GonzalezDelgado:2005}. For older ages, the SSPs are from 
\citet{Vazdekis:2015} based on BaSTi isochrones. The Z range covers eight metallicities 
(log Z/Z$_{\odot}$ = -2.28, -1.79, -1.26, -0.66, -0.35, -0.06, 0.25 and +0.40). The age is 
sampled by 37 SSPs per metallicity covering from 1 Myr to 14 Gyr. The initial mass 
function (IMF) is Salpeter. Dust effects are modeled as a foreground screen with 
a \citet{Cardelli:1989} reddening law with R$_V$ = 3.1.

%%% Subsection %%%
\subsection{Stellar masses}

The galaxy stellar masses (\mstar) used in this work are obtained by adding the masses of each 
individual spatial zone. This method takes into account variations in the stellar extinction 
and other stellar population properties, something that cannot be done with the integrated 
spectra. Masked areas due to foreground and background objects are filled in {\scshape pycasso} 
using the \nmustar\ (\mustar) radial profile \cite[see][]{GonzalezDelgado:2014a}.

% ---------------------------------------------------------------------------
\begin{table}
\small
\begin{tabular}{ccccccccc}
\hline\hline
\lmstar & E & S0 & Sa & Sb & Sbc & Sc & Sd & Total \\
\hline
8.4 - 9.9 & 12 & 1 & 1 & 6 & 4 & 44 & 42 & 110 \\
9.9 - 10.4 & 5 & 6 & 7 & 11 & 21 & 40 & 20 & 110 \\
10.4 - 10.8 & 12 & 11 & 16 & 22 & 29 & 18 & 3 & 111 \\
10.8 - 11.1 & 14 & 24 & 17 & 27 & 23 & 4 & 1 & 110 \\
11.1 - 11.3 & 21 & 22 & 31 & 20 & 14 & 2 & 0 & 110 \\
11.3 - 12.0 & 56 & 15 & 21 & 14 & 2 & 2 & 0 & 110 \\
\hline
Total & 120 & 78 & 94 & 100 & 93 & 110 & 66 & \textbf{\ngal} \\
\hline
\end{tabular}
\caption{Number of galaxies in each Hubble type and mass interval.}
\label{Tab:masshub}
\end{table}
% ---------------------------------------------------------------------------

One of our chief aims is to derive the mass growth (MG) as a function of time and explore how 
these histories relate to other parameters, in particular Hubble type, stellar mass (\mstar), 
and \nmustar\ (\mustar). Table \ref{Tab:masshub} shows the distribution of 
galaxies by Hubble type and mass bins. 
The most populated Hubble type bins are E, Sb and Sc galaxies, while the Sd bin contains  
the fewest objects (66). We have computed the sextiles (6-quantiles) of the stellar 
mass range of the whole sample in order to divide the sample into six bins (of stellar mass), each
with almost the same number of objects. 

The upper panel of Fig. \ref{Fig:massmcorsdhub} shows the 
distribution of \mstar\ as a function of Hubble type. As seen from the median value of each bin 
(inner dot in the boxplot inside the violin plots), mass correlates with Hubble type. 
Early-type galaxies (E, S0, and Sa) on average have \mstar\ $\geq$ 10$^{11}$ \msun\ while 
latter spirals like Sd have \mstar $\leq$ 10$^{9.6}$ \msun. However, the distribution of some 
Hubble types shows a significant coverage in mass. Ellipticals (E), although concentrated in 
the high-mass end as seen from the shape of the violin plot, have a small tail all the way 
down to 10$^{9}$ \mstar. This is the result of an ancillary project which was looking to populate 
this Hubble type and mass regime. To a lesser extent, late spirals (Sc, Sd) also show a 
broader distribution than other types as seen from their slightly larger interquartile ranges. 
These late spirals, 54\% of them coming from ancillary projects (59 out of 110), populate the 
low-mass end below the completeness limit of the CALIFA survey\footnote{The limits 
of the mother sample provided by \citet{Walcher:2014} are scaled to Salpeter IMF.}
(9.9 $\leq$ log \mstar (\msun) $\leq$ 11.7).

%%% Subsection %%%
\subsection{Stellar mass surface density\label{Sec:mustar}}

The total stellar mass is an important parameter in star formation and chemical enrichment 
in spheroids. However, it is in disks where \nmustar\ comes into play, 
as seen by the local relations between \mustar\ and mean metallicities 
\citep{GonzalezDelgado:2014b} and between \mustar\ and stellar ages 
\citep{GonzalezDelgado:2014a}. 

In a previous work using 107 galaxies from the CALIFA survey, \citet{GonzalezDelgado:2014a} 
showed that several galaxy-averaged properties, in particular \mustar, correlate nicely 
with the mean value over a ring at 1 $\pm$ 0.1 \hlr\ from the nucleus. 
We have estimated the galaxy-averaged \mustargal\ in the same 
way, dividing the total mass (summing the total area of each spaxel, A$_{xy}$) by the total 
area of these spaxels:

\begin{equation}
\Sigma_{\star}^{galaxy} = \frac{\Sigma_{xy} M_{xy}}{\Sigma_{xy} A_{xy}}
.\end{equation}

\begin{figure}
\centering
\resizebox{\hsize}{!}{\includegraphics{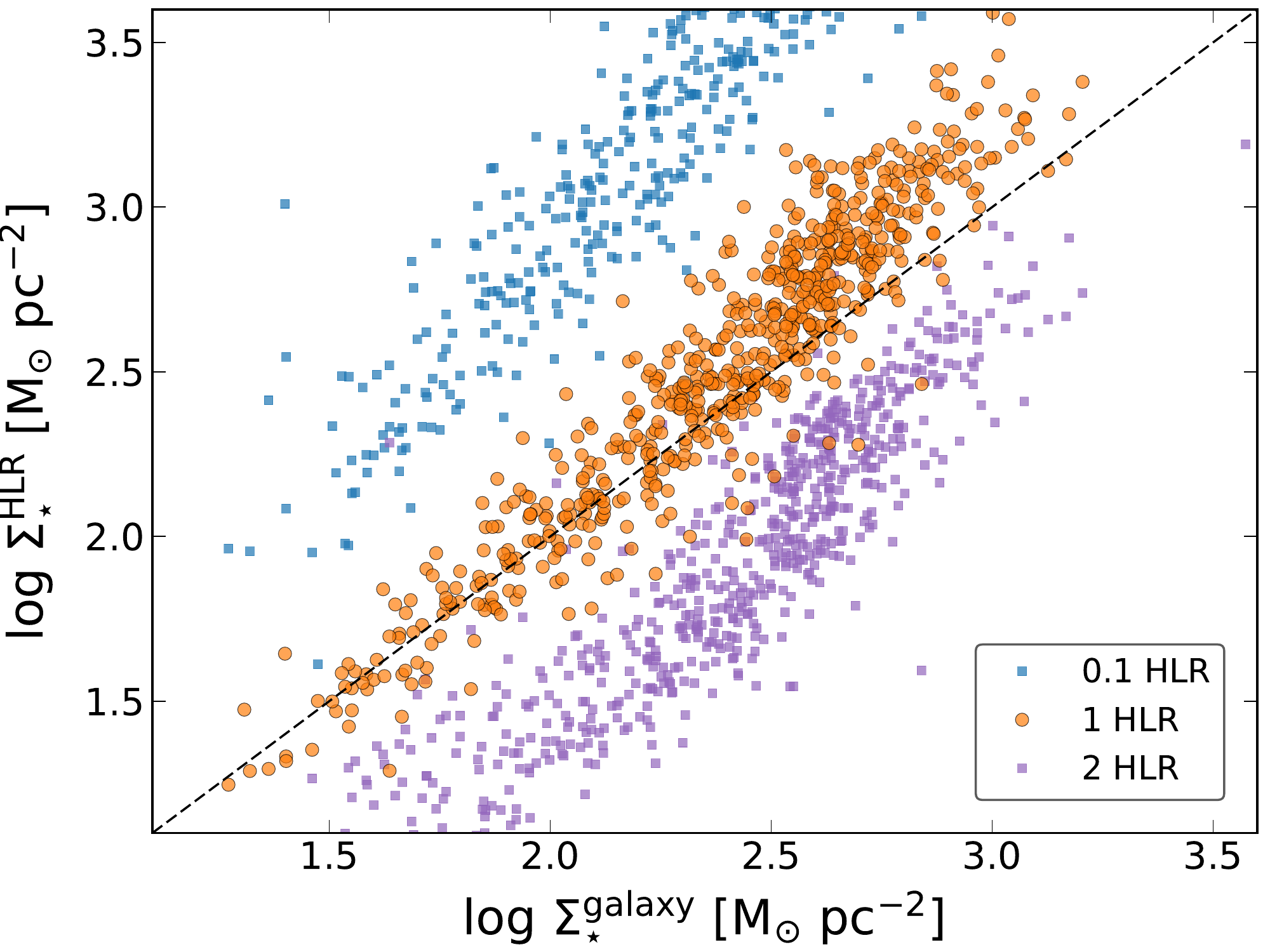}}
\caption{Relation between log \mustar$^{galaxy}$ and the \nmustar\ mean value over a ring of 
0.1 \hlr\ width at 0.1 \hlr, 1 \hlr, and 2 \hlr\ from the nucleus. The dashed black line shows 
the one-to-one relation.}
\label{Fig:mcorsdhlr}
\end{figure}

Figure \ref{Fig:mcorsdhlr} shows this relation for \ngal\ galaxies. The local value at 1 
\hlr\ follows nicely the one-to-one line except for the upper end of \mustargal, where 
\mustar$^{galaxy}$ is on-average lower than \mustarhlr\ and it correlates better at 
1.1 - 1.2 \hlr. 

%----------------------------------------------------------------- 
%-----------------------------------------------------------------

From now on, we will use the local value of \mustar\ at 1 \hlr\ (\mustar $\equiv$ \mustar$^{1 \hlr}$). 
In the same way as with the total stellar mass \mstar, we have computed the sextiles of \mustar\ 
in order to divide the sample into six equally populated bins, as shown in Table \ref{Tab:massmcorsd}. 

The middle panel of Fig. \ref{Fig:massmcorsdhub} shows the distribution of \mstar\ as a function of 
our \mustar\ bins. For the two highest \mustar\ bins (2.8 to 3.7 \mustarunits), the median 
stellar mass is nearly the same, indicative of fairly compact mass distribution, with their 
interquartile range between 10$^{10.8}$ \msun\ and 10$^{11.3}$ \msun. The rest of the \mustar\ bins 
show a more spread distribution in mass, with a correlation for their median values between \msun\ 
and \mustar, that mimics that of the total mass and Hubble type distribution 
(upper panel, Fig. \ref{Fig:massmcorsdhub}).

The distribution of \mustar\ as a function of Hubble type is shown in the bottom panel of Fig. 
\ref{Fig:massmcorsdhub}. At first sight it resembles that of \mstar. Despite ellipticals 
spanning over a large \mustar\ range, most of them peak close to the high end $\sim$ 2.8 \mustarunits. 
However, a clear difference with the mass distribution is the turning point in the 
correlation for S0 galaxies. These galaxies peak at a higher \mustar\ value close to 
3 \mustarunits, while for the mass, we have seen that Sa galaxies are the ones that, on average, 
have slightly larger values. The remaining spirals follow a similar rank relation to that of the 
mass (left panel).

% ---------------------------------------------------------------------------
\begin{table}
\small
\begin{tabular}{cccccccc}
\hline\hline
\lmstar & \multicolumn{6}{c}{\lmustar} & Total\\
&  1.2 & 2.0 & 2.4 & 2.6 & 2.8 & 3.0 &  \\
&  2.0 & 2.4 & 2.6 & 2.8 & 3.0 & 3.7 & \\
\hline
8.4 - 9.9 & 69 & 31 & 6 & 2 & 2 & 0 & 110 \\
9.9 - 10.4 & 36 & 35 & 25 & 4 & 4 & 6 & 110 \\
10.4 - 10.8 & 3 & 29 & 24 & 24 & 15 & 16 & 111 \\
10.8 - 11.1 & 2 & 8 & 26 & 24 & 25 & 25 & 110 \\
11.1 - 11.3 & 0 & 6 & 16 & 26 & 33 & 29 & 110 \\
11.3 - 12.0 & 0 & 2 & 13 & 30 & 31 & 34 & 110 \\
\hline
Total & 110 & 110 & 111 & 110 & 110 & 110 & \textbf{\ngal} \\
\hline
\end{tabular}
\caption{Number of galaxies in each total mass and \nmustar\ interval.}
\label{Tab:massmcorsd}
\end{table}
% ---------------------------------------------------------------------------

%%%%%%%%%%%%%%%%%%%%%%%%%%%%%%%%%%%%
%%%%%%%%%%%%% Section %%%%%%%%%%%%%%
%%%%%%%%%%%%%%%%%%%%%%%%%%%%%%%%%%%%
\section{Spatially resolved mass-growth histories\label{Sec:mgh}}

%%% Subsection %%%
\subsection{Radial stellar mass-growth 2D maps}

%----------------------------------------------------------------- 
\begin{figure*}
\centering
\resizebox{\hsize}{!}{\includegraphics{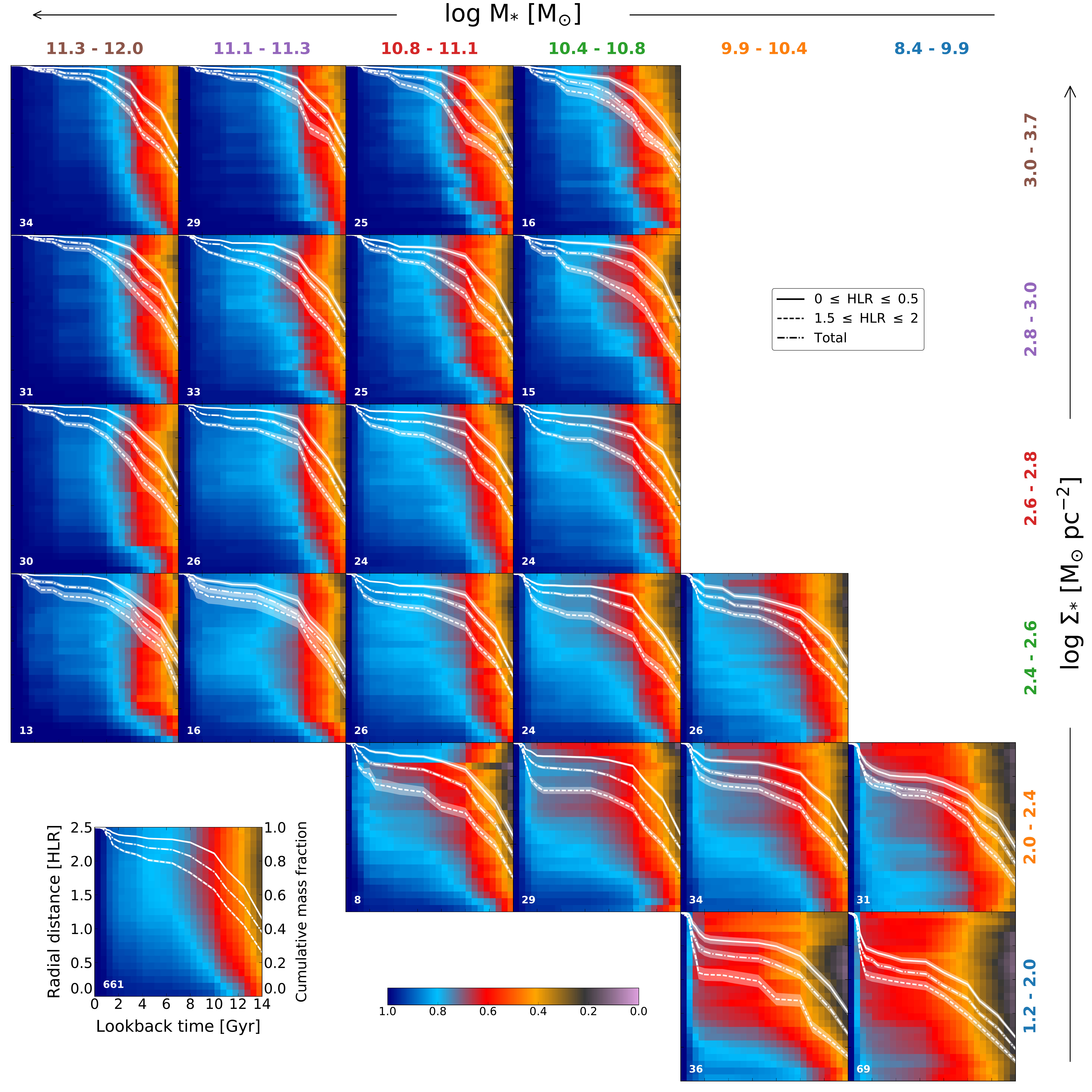}}
\caption{Radial stellar mass growth maps up to 2.5 \hlr\ stacked by stellar mass and \nmustar. 
Each radial bin is normalized to its final \mstar. Lines overplotted on the 2D maps represent the relative 
stellar mass growth of three main spatial regions (white): inner 0.5 \hlr\ (solid line), 
1.5 $\leq$ \hlr\ $\leq$ 2 (dashed line), 
and total (dash-dotted line). The (white) shaded areas represent the standard deviation of the mean. 
The lower left number in each map indicates the number of galaxies stacked for that particular 
bin. The lower-left inset box shows the results for all \ngal\ galaxies. Mass and \nmustar\ bins grow from 
right to left and bottom to top, respectively.}
\label{Fig:mg2D}
\end{figure*}
%-----------------------------------------------------------------

Following the space $\times$ time diagrams (R $\times$ $t$) introduced by \citet{CidFernandes:2013} 
(see their Fig. 12), we apply the same representation scheme to the mass growth of galaxies. These 
diagrams show in the x-axis the lookback time, while the y-axis is the radial distance, compressing 
the $xy$ information into $R$, producing radially averaged maps of the desired quantity. 

%----------------------------------------------------------------- 
\begin{figure*}
\centering
\resizebox{\hsize}{!}{\includegraphics{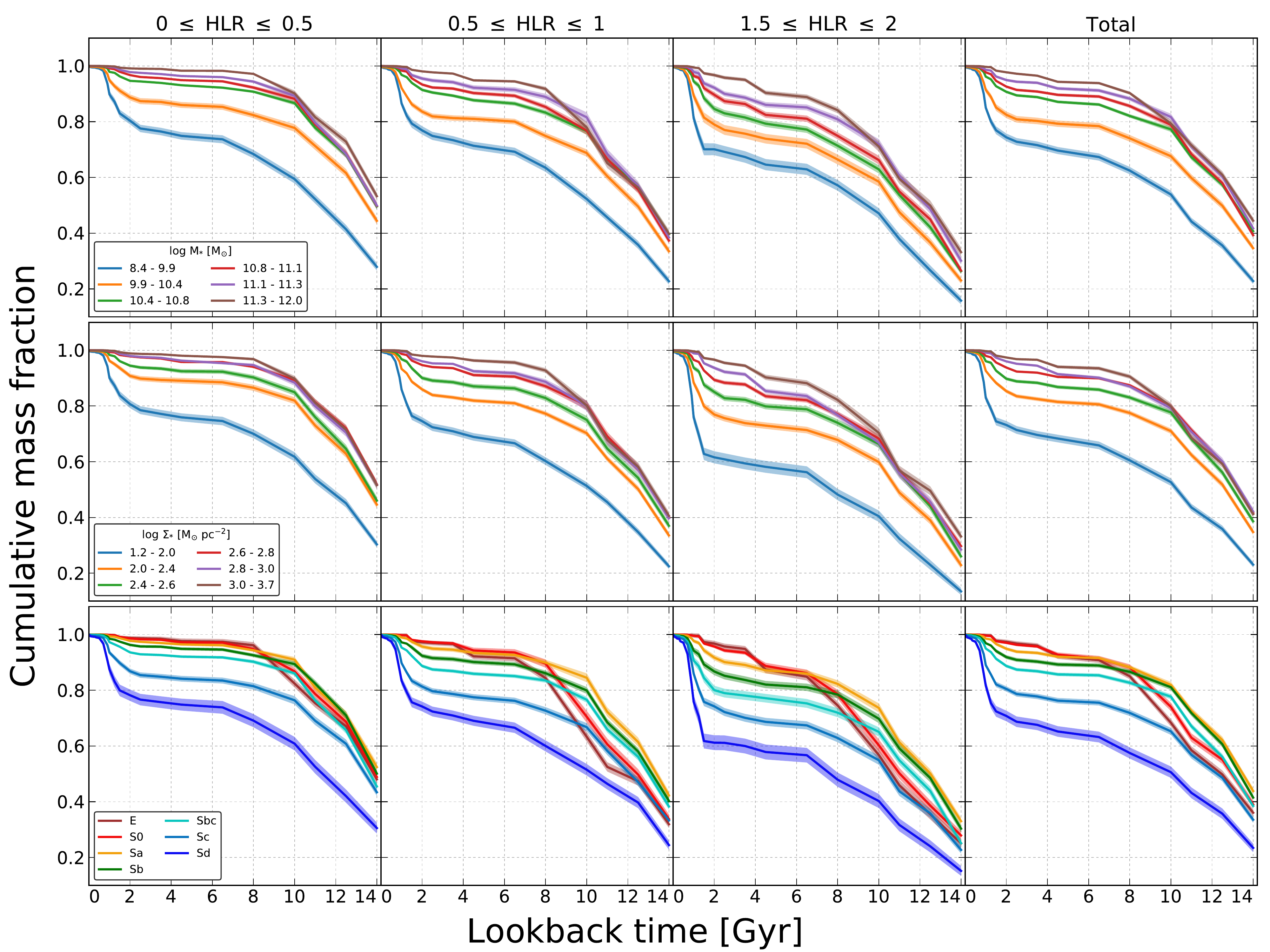}}
\caption{Relative stellar mass growth of three spatial regions ($\leq$ 0.5 \hlr, 0.5 $\leq$ \hlr\ $\leq$ 1, 
1.5 $\leq$ \hlr\ $\leq$ 2) and total, binned by stellar mass (upper panel), \nmustar\ (middle panel) 
and Hubble type (lower panel).}
\label{Fig:mgTotal}
\end{figure*}
%-----------------------------------------------------------------

Figure \ref{Fig:mg2D} shows the R $\times$ $t$ maps of the mass assembly, a means to represent the 
mass growth history of galaxies in 2D. \stl\ provides a wealth of information for 
each single spectrum, in particular the mass in stars for each mass fraction vector $m_{xytZ}$, 
derived by \stl\ from the population vector taking into account the stellar extinction 
and M/L spatial variation. This vector is collapsed in the metallicity (Z) axis, and then the cumulative 
mass function along the time axis is computed for each spatial location. The cumulative function 
for a given spaxel is calculated by adding all mass formed up to a given lookback time and dividing 
by the final stellar mass. In order to stack any given property, galaxy by galaxy, we express the distance 
in a common metric such as the half light radius (HLR), the semi-major axis 
length of the elliptical aperture which contains half of the total light of the galaxy at the rest-frame 
wavelength 5635 \AA. Radial apertures of 0.1 \hlr\ in width were used to extract the radial profiles.

The averaged mass growth 2D maps of \ngal\ galaxies in 6 \mstar\ bins and 6 \mustar\ bins is 
shown in Fig. \ref{Fig:mg2D}. The maps contain also the relative stellar mass growth curve of three 
spatial regions: inner 0.5 \hlr\ (solid line), 1.5 $\leq$ \hlr\ $\leq$ 2 (dashed line), and total 
(dash-dotted line). The mass growth curve for every component and galaxy is computed by averaging the 
spatial locations within the required distance and then the individual value is stacked for all galaxies. 
The standard error of the mean is estimated at every radial distance and for each spatial region. We 
show only averaged quantities with a minimum number of seven galaxies per bin ($\sim$ 1\% of the total 
sample). Both the 2D maps and spatial component curves show clearly the inside-out assembly of the 
stellar mass, which is preserved at any \mstar\ and \mustar\ bin and time, even for the low-mass end. 
The assembly of the mass in the outskirts of the galaxies is more extended in time than the 
inner 0.5 \hlr. The averaged growth curve for the whole galaxy always keeps in an intermediate loci  
between inner and outer regions.

%%% Subsection %%%
\subsection{Spatial averaged mass assembly}

We can collapse the results of Fig. \ref{Fig:mg2D} in both axes and stack also by Hubble type in order to 
assess the global shape of the mass assembly curves. Figure \ref{Fig:mgTotal} shows the averaged results for 
three different spatial regions ($\leq$ 0.5 \hlr, 0.5 $\leq$ \hlr\ $\leq$ 1, 1.5 $\leq$ \hlr\ $\leq$ 2) and 
for the whole galaxy. We can see at work a stellar-mass ranking (upper panel) in the global curve, that is, 
archaeological downsizing\footnote{See \citet{Fontanot:2009} for a summary of different types of downsizing.},  
indicating that stars in more massive galaxies formed at an earlier epoch and in a shorter time span, 
confirming our previous result from the analysis of a sample of 105 galaxies \citep{Perez:2013}. The downsizing 
effect is also spatially preserved for different regions. The growth curve from the low to medium mass regime 
are well segregated, while the curves for the bins up to the higher-mass end display tighter ranking. It is 
worth noting that the first two bins cover two order of magnitudes 
in mass (10$^{8.4}$ to 10$^{10.4}$ \msun) while the four upper mass bins cover a smaller range from 
(10$^{10.4}$ to 10$^{12.0}$ \msun). By construction all bins contain the same number of galaxies.

The anti-correlation between the \nmustar\ of galaxies and the formation epoch of the 
stars is preserved. Again, the first two lower bins span for over two orders of magnitude in \mustar\  
(10$^{1.2}$ and 10$^{2.4}$), while the rest are concentrated between 10$^{2.4}$ and 10$^{3.7}$ 
\mustarunits\, displaying a tighter ranking of the mass growth curves.

%----------------------------------------------------------------- 
\begin{figure*}
\centering
\resizebox{\hsize}{!}{\includegraphics{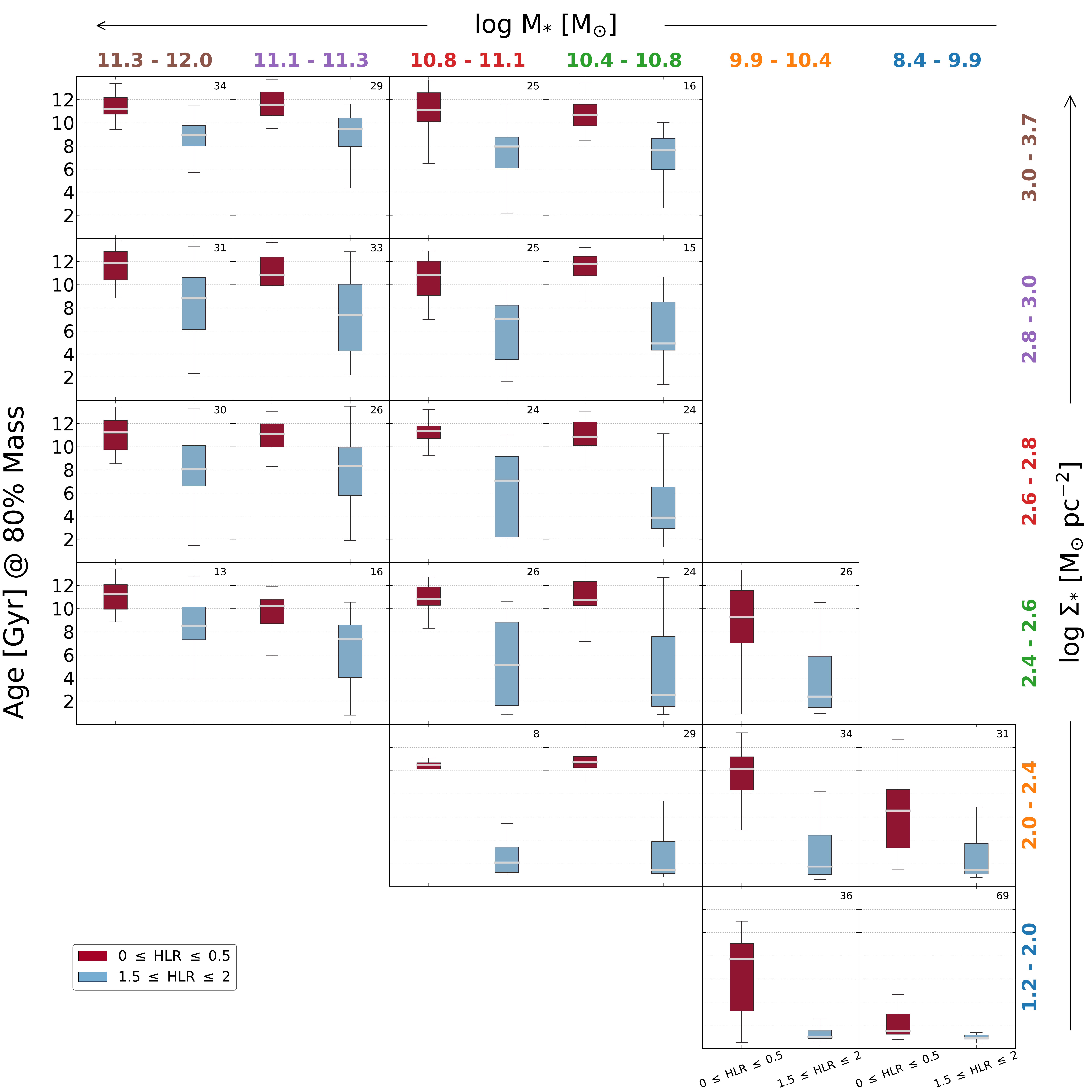}}
\caption{Box plot diagram of the age (in Gyr) at which the inner ($\leq$ 0.5 \hlr) and outer (1.5 $\leq$ \hlr\ $\leq$ 
2) spatial regions grow to 80\% of their final stellar mass stacked by stellar mass and \nmustar. The gray horizontal 
line in each box plot shows the median of the distribution. The upper right number in each panel indicates the 
number of galaxies stacked for that particular bin.}
\label{Fig:mgAp}
\end{figure*}
%-----------------------------------------------------------------

The cumulative mass fraction plots show that both \mstar\ and \mustar\ are fundamental parameters segregating 
the mass assembly curves. However, this is not the case for all morphological types, where the downsizing does 
not hold for early types (E and S0) at all epochs, albeit the ranking holds for the remaining Hubble types. 
On the one hand, we have seen that ellipticals span a significantly large mass range (upper panel, 
Fig. \ref{Fig:massmcorsdhub}) and the median mass of Sa galaxies is slightly larger than S0 galaxies. On the 
other hand, S0 galaxies have, on average, largest \nmustar\ values (bottom panel, Fig. 
\ref{Fig:massmcorsdhub}) and also show a fairly broad \mustar\ range. The combination of these factors 
seems to be reason for the break of the ``top-down'' mass assembly behavior for E and S0 galaxies. 

%%% Subsection %%%
\subsection{Characteristic mass assembly epoch}

%----------------------------------------------------------------- 
\begin{figure*}
\centering
\resizebox{\hsize}{!}{\includegraphics{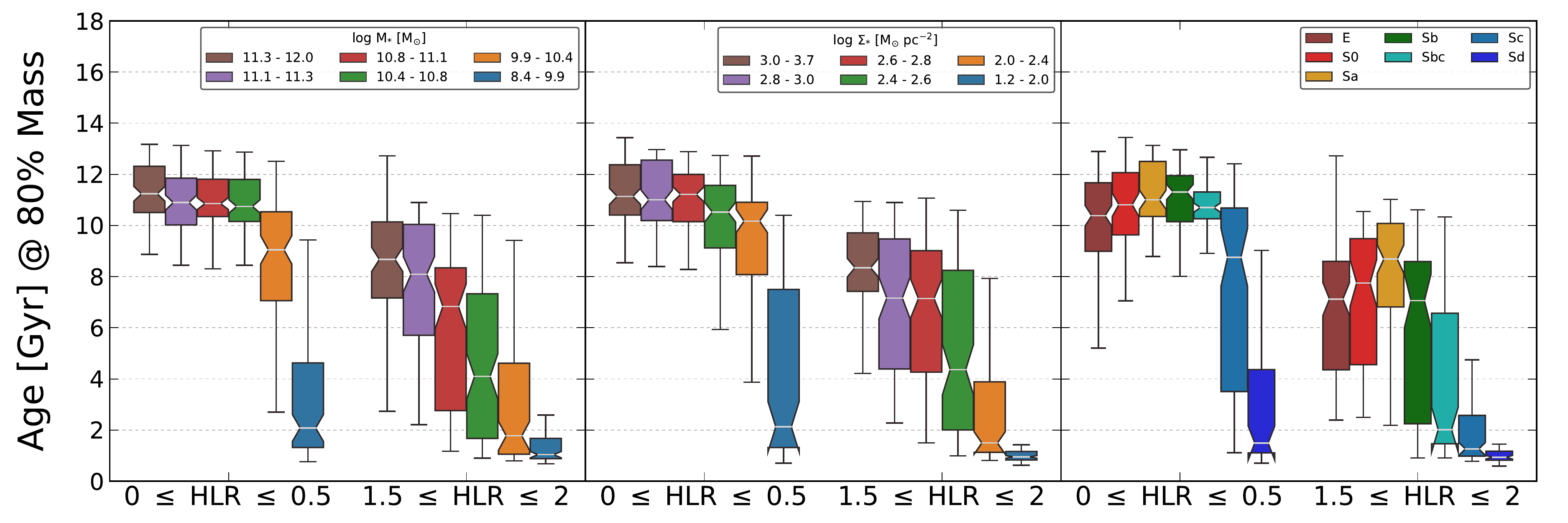}}
\caption{Same as Fig. \ref{Fig:mgAp} but binning only in one property: stellar mass (left),  
\nmustar\ (center), and Hubble type (right).}
\label{Fig:mgApTotal}
\end{figure*}
%-----------------------------------------------------------------

For all stellar mass and \nmustar\ bins, the average 2D maps show a clear inside-out mass growth. To better quantify 
the rate at which the galaxies have grown their mass, 
we compute the epoch where 80\% of the total assembled mass (\tg{80}) for each spatial region was built. We choose two 
distinct regions: $\leq$ 0.5 \hlr\ representative of the inner part and 1.5 $\leq$ \hlr\ $\leq$ 2 for the outer 
component.
Figure \ref{Fig:mgAp} displays the distribution of stellar mass assembly ages in bins of \mstar\ and \mustar. 
We have not stacked the individual values in each bin, to better visualize the differences and scatter of the 
samples by means of box plots. The interquartile range shows a clear difference: outer regions achieve their  
80\% stage at a more recent epoch that inner regions. In most cases, the interquartile ranges are well 
separated for both regions, an indication of two different assembly ages. Also, the assembly epoch for the outer 
regions presents more scatter in the intermediate \mstar\ and \mustar\ bins. The median values (gray line inside the 
boxplots) for the ages of the inner region populations are systematically older than the outer ones; a signal that 
the ``downsizing'' effect is spatially preserved. The largest difference for the median values of \tg{80} between 
inner and outer regions is found for bin \mmbin{10.4}{10.8}{2.0}{2.4}, where the inner region formed 80\% of its 
mass 9.2 Gyr before the outer region, followed by and \mmbin{10.8}{11.1}{2.0}{2.4} (8.6 Gyr) and 
\mmbin{9.9}{10.4}{2.0}{2.4} (8.6 Gyr). It is worth noting all three cases belong to the same \mdbin{2.0}{2.4} bin. 
As for the smaller differences between inner and out regions, these are found in bins \mmbin{8.4}{9.9}{1.2}{2.0} 
(0.5 Gyr) and \mmbin{8.4}{9.9}{2.0}{2.4} (1.7 Gyr), at the lowest \mstar\ and \mustar\ ranges.

For a quantitative proof of the inside-out mass growth mechanism, we conducted statistical tests for comparing 
these two populations medians. The Paired t-Test is valid when the number of pairs is $>$ 30, or when the 
\textit{differences} of the pairs follow a normal distribution (for number of pairs $<$ 30), and it does not 
assume that the groups are homoscedastic. In case these assumptions are not met, a non-parametric paired difference 
test can be used, like the Wilcoxon signed-rank test. We have conducted both tests for all samples in Fig. 
\ref{Fig:mgAp} and both conclude that the paired population means are not equal, and therefore, the assembly ages 
for the inner regions are larger than the outer regions for \textit{all} \mstar\ and \mustar\ bins.

As we did for the mass assembly curves, we can collapse Fig. \ref{Fig:mgAp} in both axes. Figure \ref{Fig:mgApTotal} 
shows the results comparing the characteristic mass assembly at \tg{80} collapsing by \mstar, \mustar\ and Hubble 
type for the ($\leq$ 0.5 \hlr) and outer (1.5 $\leq$ \hlr\ $\leq$ 2) regions. Each bin now includes all \ngal\ 
galaxies. Again, the median values of the outer regions is always below that of the inner region, for every 
\mstar, \mustar, and Hubble type bin, as is also the case for most of the interquartile range; an indication that, 
on average, the inner parts form earlier than the outer parts of the galaxy. The spread in the values of \tg{80}, 
judged from the interquartile range in the box plots, is, in general, larger in the outer regions. This is also
particularly evident  from the relative wider interquartile range for the \mbin{9.9}{10.4} bin and \mdbin{2.0}{2.4} 
and \mdbin{1.2}{2.0} bins of the inner regions (left panel). 

The median \tg{80} epoch for the inner regions is very similar for the four higher-mass bins (left panel, 
10.4 to 12.0 log \mstar), with values of 11.2 Gyr for the higher-mass bin and $\sim$ 10.7 Gyr for the other 
three. A similar behavior is displayed in the \mustar\ bins (middle panel). The outer region, both for 
\mstar\ and \mustar, shows a more pronounced top-down sequence. As for the Hubble type (right panel), 
the ranking for the median values is not preserved either by the inner or outer regions. This trend is 
understandable in the light of Fig. \ref{Fig:massmcorsdhub}, where we see a similar pattern in the 
\mustar\ distribution as a function of Hubble type.

%%%%%%%%%%%%%%%%%%%%%%%%%%%%%%%%%%%%
%%%%%%%%%%%%% Section %%%%%%%%%%%%%%
%%%%%%%%%%%%%%%%%%%%%%%%%%%%%%%%%%%%

%%% Subsection %%%
\section{Half-mass versus half-light radius\label{Sec:hmr_hlr}}

%----------------------------------------------------------------- 
\begin{figure*}
\centering
\resizebox{\hsize}{!}{\includegraphics{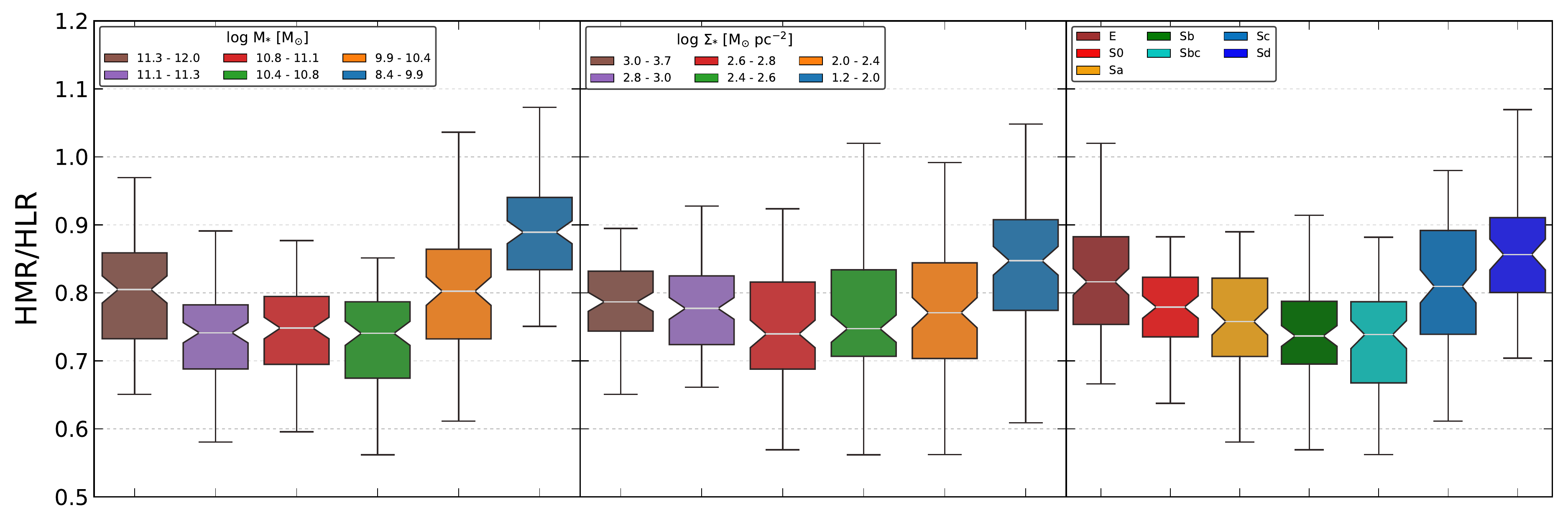}}
\caption{Box plot diagram of the ratio between the half-mass and the half-light radius binned by Hubble type (left), 
stellar mass (center), and \nmustar\ (right). The gray horizontal line in each notched box plot shows the 
median of the distribution.}
\label{Fig:hmr_hlr}
\end{figure*}
%-----------------------------------------------------------------

So far we have used a galaxy metric based on their light, the \hlr. However, stellar mass is a key
parameter in describing the mass assembly, and therefore the half mass radius (\hmr) seems a 
more suitable physical metric for galaxy evolution. The relation between these two metrics depends 
on the variations of the SFH and/or extinction, which produce a spatially dependent M/L ratio.

Thanks to our spatially resolved SFH, we can obtain the relation between \hlr\ and \hmr. We estimate the 
latter using the 2D distribution of \mustar, defined as the semi-major axis length of the elliptical 
aperture which contains half of the total mass of the galaxy. 
Figure \ref{Fig:hmr_hlr} shows their relation as a function of \mstar, \mustar, and Hubble type. 
The median half-mass size is always smaller than half-light radii for all three parameters. 
The lowest ratio is found in three of the most massive bins (10.4 to 11.1 log \mstar), where 
galaxies are 25\% smaller in mass than in light, very similar to the value found by \citet{Szomoru:2013} 
for galaxies at  0.5 $<$ $z$ $<$ 2.5 with > 10$^{10.7}$ \msun. Low-mass galaxies \mbin{8.4}{9.9}  
show the largest ratio with \hmr/\hlr\ $\sim$ 0.89. The trend is also followed by \mustar, 
although the relation is less pronounced. As for the Hubble type, Sb and Sbc galaxies present 
the lowest \hmr/\hlr\ ratio (0.74), in accordance with our previous results 
\citep{GonzalezDelgado:2015}. 

The \hlr\ was estimated in the observed light profiles. In previous works, we have presented detailed 
analysis of the several stellar profiles of several stellar properties, including the extinction 
\citep[see][]{GonzalezDelgado:2015}. Thanks to the 2D extinction maps we can explore whether dust 
correction could flatten the \hmr/\hlr. We corrected the light profiles using the extinction maps
and estimated the \hlr$^{DeRed}$. Extinction has a mild effect on the flattening of the 
\hmr/\hlr$^{DeRed}$. For the low-mass galaxies bin, the ratio flattens by 1\%, while for the range 
\mbin{9.9}{11.3} the increase in the ratio is between 5\% and 7\%, being close to 3\% for the 
most massive galaxies.

%%%%%%%%%%%%%%%%%%%%%%%%%%%%%%%%%%%%
%%%%%%%%%%%%% Section %%%%%%%%%%%%%%
%%%%%%%%%%%%%%%%%%%%%%%%%%%%%%%%%%%%
\section{Mass-weighted age profiles\label{Sec:mage}}

Following our previous definitions \citep{CidFernandes:2013}, the mass-weighted mean log stellar 
age is defined as 

\begin{equation}
\langle \log\, age \rangle_{M} = \sum_{t,Z} m_{tZ} \times \log\, t
,\end{equation}

\noindent where $m_{tZ}$ is the fraction of mass attributed to the base element with age $t$ and metallicity $Z$.

\begin{figure*}
\centering
\resizebox{\hsize}{!}{\includegraphics{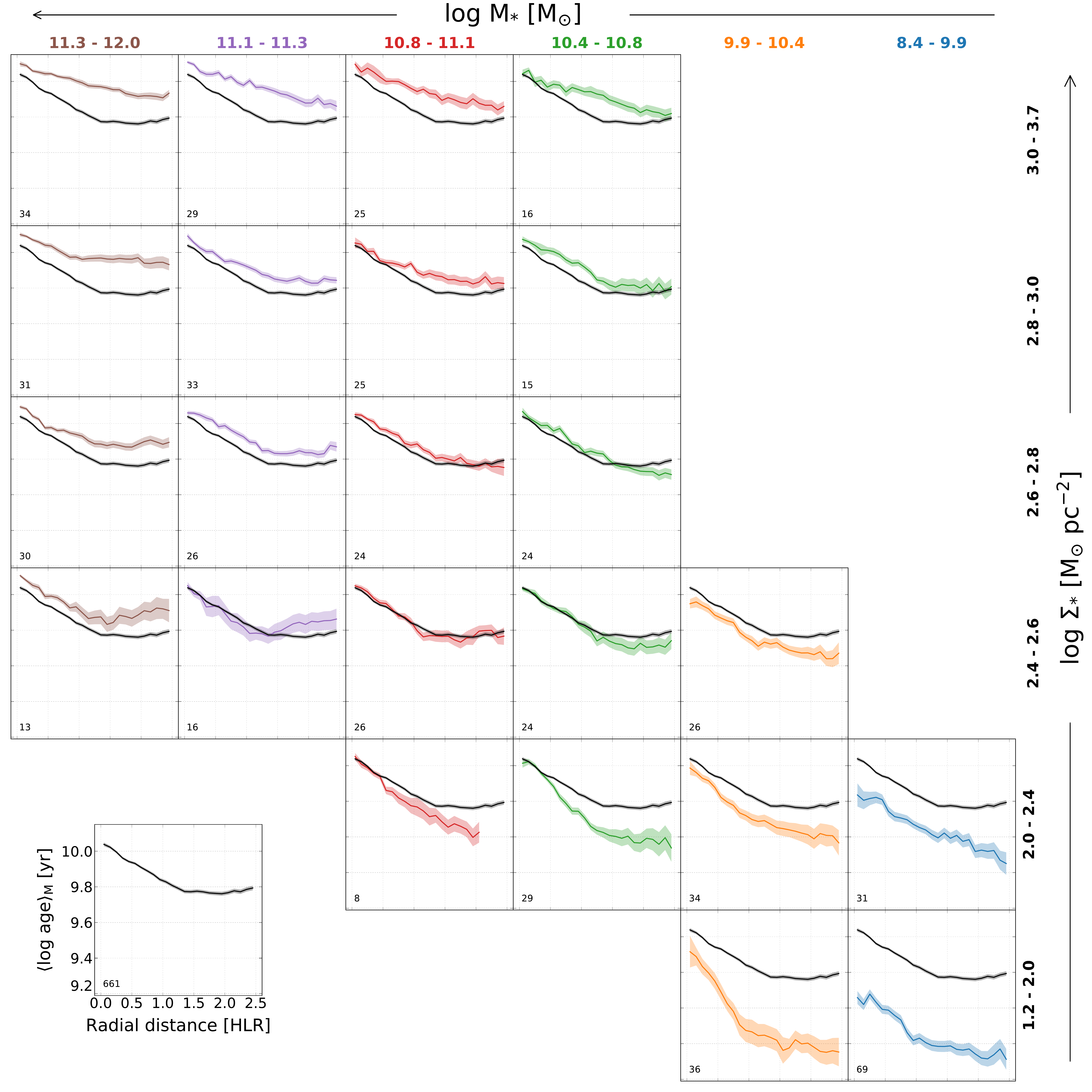}}
\caption{Radial profiles (in units of \hlr) of the mass-weighted age stacked by stellar mass and \nmustar. The 
gray line shows the median profile obtained with the \ngal\ galaxies. This profile is also shown in the lower-left 
inset box. The number of galaxies in each bin is indicated in the lower-left side of each panel.}
\label{Fig:at_mass__r}
\end{figure*}

In \citet{GonzalezDelgado:2014a} we analyzed the radial profiles of the luminosity-weighted log stellar age \agel\
as a function of mass and Hubble type and \agem\ as a function of Hubble type for 300 galaxies. 
Figure \ref{Fig:at_mass__r} goes a step further and explores the \agem\ radial profiles \textit{both} 
as a function of \mstar\ and \mustar for \ngal\ galaxies. Because of the weight of old populations, \agem\ spans 
a smaller dynamical range than light-weighted age. However, \agem\ and its gradient is less affected by young 
populations whose contribution is negligible to the mass growth. Negative gradients are detected in all 
galaxies. The radial profiles of \agem\ decrease outward for all bins, which are more pronounced for low-\mstar\ 
and low-\mustar\ galaxies, a clear effect of the inside-out growth of galaxies. The \nmustar\ seems 
to play a role between galaxies of the same \mstar, with a general downsizing pattern particularly notable for 
low \mstar. Most of the stellar mass in the center was formed more than 10 Gyr ago, except for the lowest-\mstar\  
and -\mustar\ bins. There is a general change in the trend around the spatial region 1 - 1.5 \hlr, where the profile 
starts to flatten, although in a less noticeable manner for low-mass galaxies. The \mmbin{10.8}{11.1}{2.4}{2.6} bin 
has a \agem\ radial profile representative of the whole sample, very close in shape and value to the global one.

%----------------------------------------------------------------- 
%-----------------------------------------------------------------

%%%%%%%%%%%%%%%%%%%%%%%%%%%%%%%%%%%%
%%%%%%%%%%%%% Section %%%%%%%%%%%%%%
%%%%%%%%%%%%%%%%%%%%%%%%%%%%%%%%%%%%
\section{Discussion\label{Sec:discussion}}

%%% Subsection %%%
\subsection{Global mass assembly time scales}

%----------------------------------------------------------------- 
\begin{figure*}
\centering
\resizebox{\hsize}{!}{\includegraphics{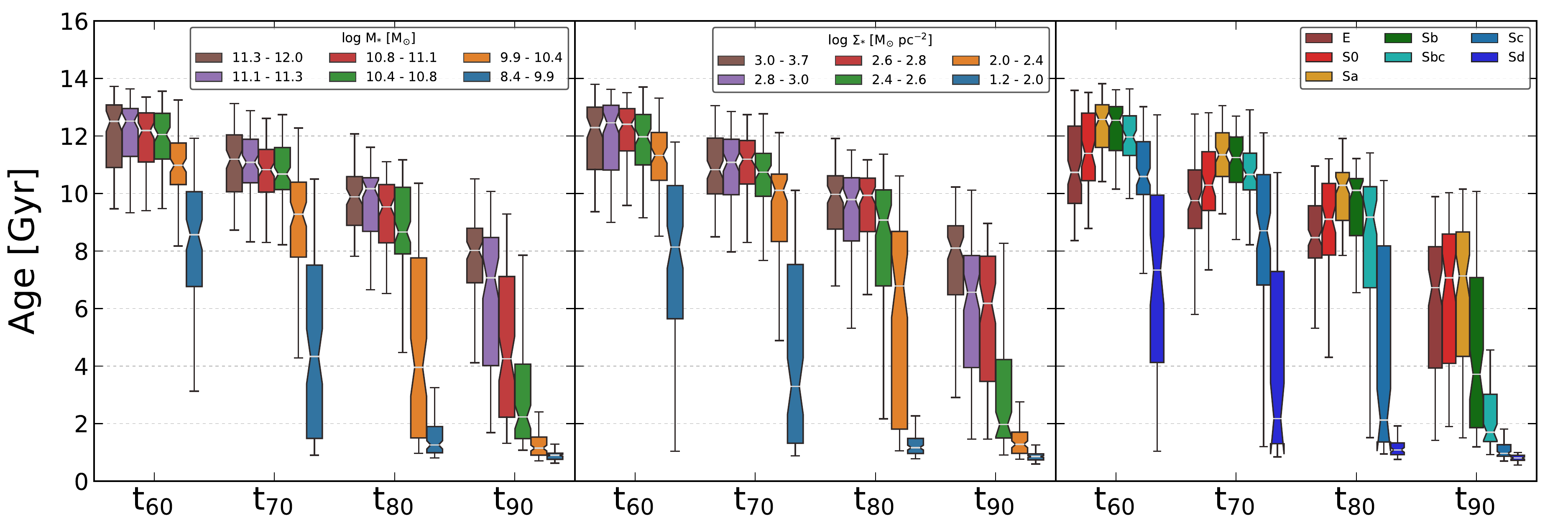}}
\caption{Box plot diagram binning by stellar mass (left), \nmustar\ (center), and Hubble type (right) of the age 
(in Gyr) at which the galaxies grow to 60\%, 70\%, 80\%, and 90\% of their final total stellar mass. The gray 
horizontal line in each notched box plot shows the median of the distribution.}
\label{Fig:mgTimeTotal}
\end{figure*}
%-----------------------------------------------------------------

The global mass assembly for the whole galaxy for different epochs is plotted in Fig. \ref{Fig:mgTimeTotal}. 
The top-down trend is nicely followed by all mass bins (left panel) at all characteristic epochs, except 
for \tg{80}, where the median of the second highest mass bin is slightly larger than the most massive bin. 
The downsizing effect for \tg{90} is the more pronounced among these four epochs, albeit with an interquartile 
range that is larger for the intermediate mass bins. The global mass assembly for the \nmustar\ (central panel) 
has a similar time scale for the highest bins, and the difference between bins increases also for \tg{90}, 
where there is a clear top-down trend for the median values. The Hubble type follows a similar pattern to 
the spatially resolved case (Fig. \ref{Fig:mgApTotal}), where the highest value is obtained by Sa galaxies. 

\begin{figure*}
\centering
\resizebox{\hsize}{!}{\includegraphics{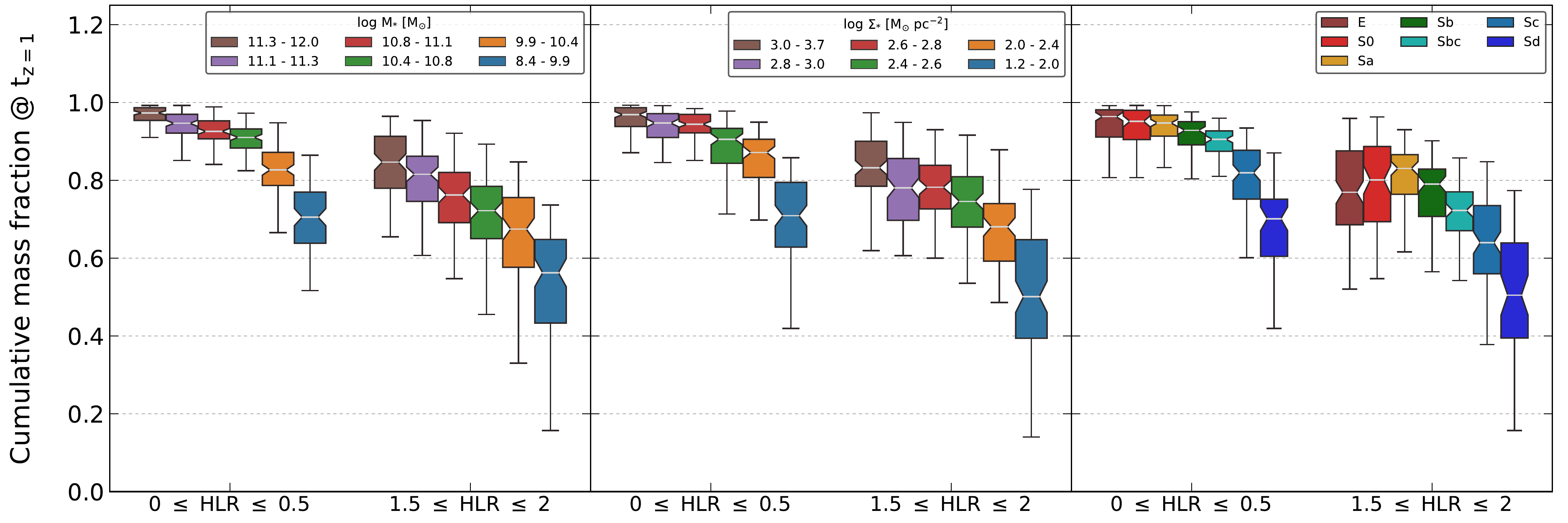}}
\caption{Box plot diagram of the relative stellar mass growth for the inner ($\leq$ 0.5 \hlr) and outer 
(1.5 $\leq$ \hlr\ $\leq$ 2) regions at $z\, =\, 1$ ($\sim$ 7.7 Gyr). The gray horizontal line in each 
notched box plot shows the median of the distribution.}
\label{Fig:mgAptz}
\end{figure*}

So far we have considered the time scales at which the galaxy or a component of the galaxies attains 
a percentage of their mass. Another way of looking at the mass assembly is by slicing at a particular 
time and comparing the fractions of masses formed by that epoch. This is presented in Fig. 
\ref{Fig:mgAptz} for $z\, =\ 1$ ($\sim$ 7.7 Gyr). Both the mass fraction of the inner ($\leq$ 0.5 \hlr) 
and outer regions (1.5 $\leq$ \hlr\ $\leq$ 2) is shown for comparison. The spatially resolved downsizing 
pattern for the inner and outer regions is preserved both for \mstar\ and \mustar. 
On average, the inner regions of the most massive galaxies (\mbin{11.3}{12.0}) have already 
attained almost all their mass at $z\, =\ 1$ ($\sim$ 97\%), while the inner regions of low-mass galaxies 
(\mbin{8.4}{9.9}) are still at $\sim$ 70\% of their final mass. The outer regions follow a similar trend, 
although as noted previously, with slightly larger scatter, increasing towards low-mass galaxies. As 
seen in Sect. \ref{Sec:multiassembly}, the assembly rates of this mass regime seem to be 
strongly dependent on other parameters such as the Hubble type, leading to a higher scatter. The outer 
region of massive galaxies is at $\sim$ 85\% of its final mass at $z\, =\ 1$, while for low-mass galaxies this is $\sim$ 56\%. The \nmustar\ mimics the spatially resolved downsizing pattern
seen for the mass and their average cumulative fraction values. The situation changes for the Hubble 
type. The top-down pattern is followed for the inner regions from E to Sd galaxies, but this trend is 
broken for the outer parts, where again the peak is found for the Sa bin. This can be understood in 
light of the SFH of E and S0 galaxies. In \citet{GonzalezDelgado:2017} we have found these galaxies to
have an important active epoch of mass assembly at $z$ $\sim$ 1 in their outer parts (1 - 1.5 \hlr).

%----------------------------------------------------------------- 
%-----------------------------------------------------------------

%%% Subsection %%%
\subsection{Multivariate mass assembly\label{Sec:multiassembly}}

%----------------------------------------------------------------- 
\begin{figure*}
\centering
%\resizebox{\hsize}{!}{\includegraphics{\sprop_growth__tg_massmcorsd_BoxPlotND_\sfig}}
%\resizebox{\hsize}{!}{\includegraphics{\sprop_growth__tg_masshub_BoxPlotND_\sfig}}
%\resizebox{0.49\textwidth}{!}{\includegraphics{\sprop_growth__tg_massmcorsd_BoxPlotND_\sfig}}
%\resizebox{0.49\textwidth}{!}{\includegraphics{\sprop_growth__tg_masshub_BoxPlotND_\sfig}}
\resizebox{\hsize}{!}{\includegraphics{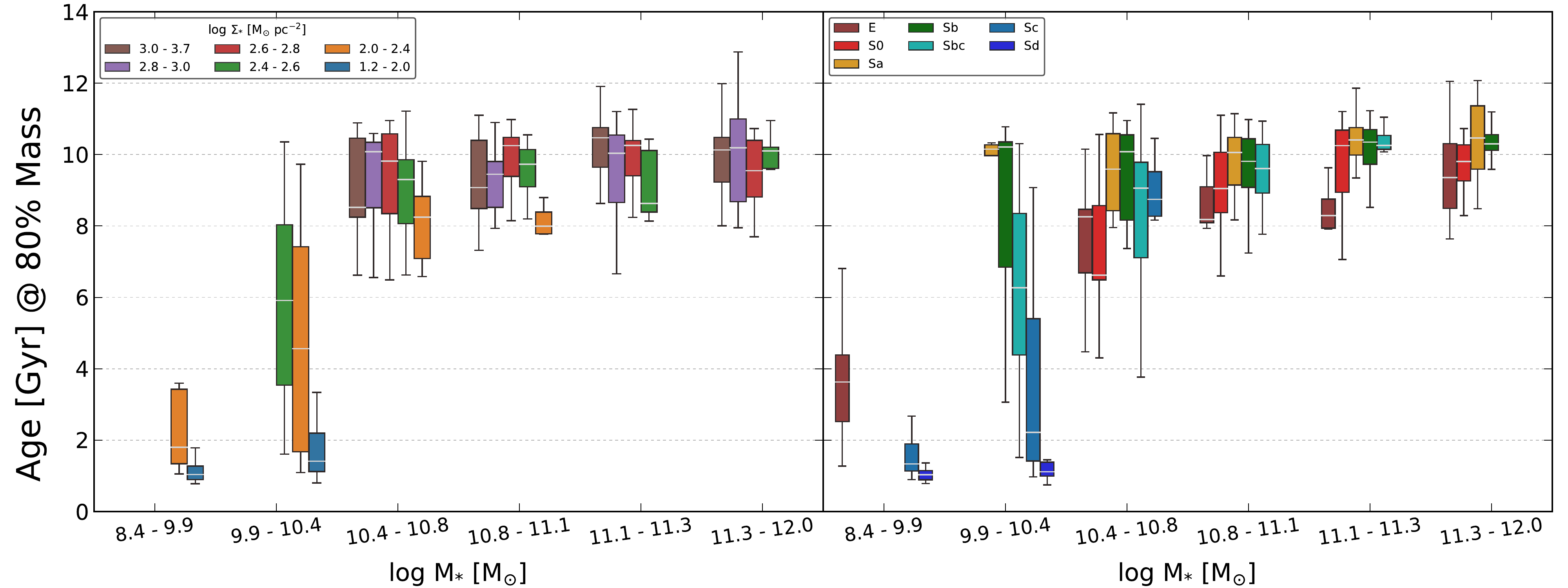}}
\caption{\textit{Left panel:} Box plot diagram of the age (Gyr) at which the galaxies grow to 80\% of their 
final stellar mass segregated by \nmustar\ bins. 
\textit{Right panel:} Box plot diagram of the age (Gyr) at which the galaxies grow to 80\% of their final stellar mass 
segregated by Hubble-type bins. The gray horizontal line in each box plot shows the median of the distribution.} 
\label{Fig:multivar_tg}
\end{figure*}
%-----------------------------------------------------------------

We have reported that the global and spatially resolved mass assembly of galaxies have a distinct behavior 
as functions of total stellar mass, \nmustar\ and Hubble type. However within a bin of a particular 
parameter, galaxies can have different properties, and there might be secondary dependencies on one (or more) 
of these properties. In fact in \citet{GonzalezDelgado:2015} we showed that radial variations in averaged 
light-weighted ages are modulated primarily by galaxy morphology, and only secondarily by the galaxy mass. 
We have seen at work in Figs. \ref{Fig:mg2D}, \ref{Fig:mgAp}, and \ref{Fig:at_mass__r} a segregation in both 
stellar mass and \nmustar. Here we want to assess whether or not there is any dependence on some of the reported 
parameters.

In the left panel of Fig. \ref{Fig:multivar_tg}, we present the distribution of the global \tg{80} both as a 
function of stellar mass and \nmustar\ bins. The number of galaxies in each bin is the same as in Table 
\ref{Tab:massmcorsd} and Figs. \ref{Fig:mg2D}, \ref{Fig:mgAp}, and \ref{Fig:at_mass__r}. For galaxies in 
the first two lowest-mass bins (log \mstar\ from 8.4 to 10.4), \nmustar\ clearly segregates 
galaxies by formation epoch \tg{80}, especially for the lowest \nmustar\ bin \mdbin{1.2}{2.0}. For 
intermediate-mass bins (log \mstar\ 10.4 to 11.1), \nmustar\ seems to play a role for the lowest-\mustar\ bin available in that stellar mass range (\mdbin{2.0}{2.4}), but higher \nmustar\ values 
appear to be insensitive in this regime and in the highest stellar mass bins. 

The right panel of Fig. \ref{Fig:multivar_tg} displays the distribution of the global \tg{80} both as a function 
of mass and Hubble type. As in previous Figures, we show only bins with a minimum number of seven galaxies (see 
Table \ref{Tab:masshub}). For galaxies in the first two lowest-mass bins (log \mstar\ from 8.4 to 10.4), 
Hubble type clearly segregates by formation epoch \tg{80}. The scatter of \mbin{9.9}{10.4} seen in the 
left panel of Fig. \ref{Fig:multivar_tg} seems here to be related to the time scales of different morphological 
types in this mass range. For intermediate- and high-mass bins, there seems to be a slight dependence for early 
types (E and S0), whereas later types reach a ``plateau'' within each mass bin.

From this latter plot it is evident that low-mass galaxies and the low end of intermediate-mass galaxies 
(log \mstar\ from 8.4 to 10.4) show a clear diversity in their assembly times, diversity that 
correlates with \nmustar\ and strongly with Hubble type. These properties still have some 
effect on their time scales for higher-mass ranges, but the differences are milder. Using 105 CALIFA 
galaxies, \citet{Perez:2013} suggested an outside-in assembly formation for lower-mass galaxies. 
As we have seen, the assembly epoch is very diverse for low-mass galaxies, and thus their low statistics 
in this mass regime could have biased this result.
\citet{IbarraMedel:2016} also found a significant diversity on the assembly epochs for low-mass galaxies, 
although they only made a distinction between early- and late-type galaxies. Figure \ref{Fig:multivar_tg} 
reveals that this diversity is clearly segregated, for log \mstar\ from 8.4 to 10.4, in terms of 
\nmustar\ and Hubble type.

%%% Subsection %%%
\subsection{Radial gradients of the mass-weighted age}

%----------------------------------------------------------------- 
\begin{figure*}
\centering
\resizebox{\hsize}{!}{\includegraphics{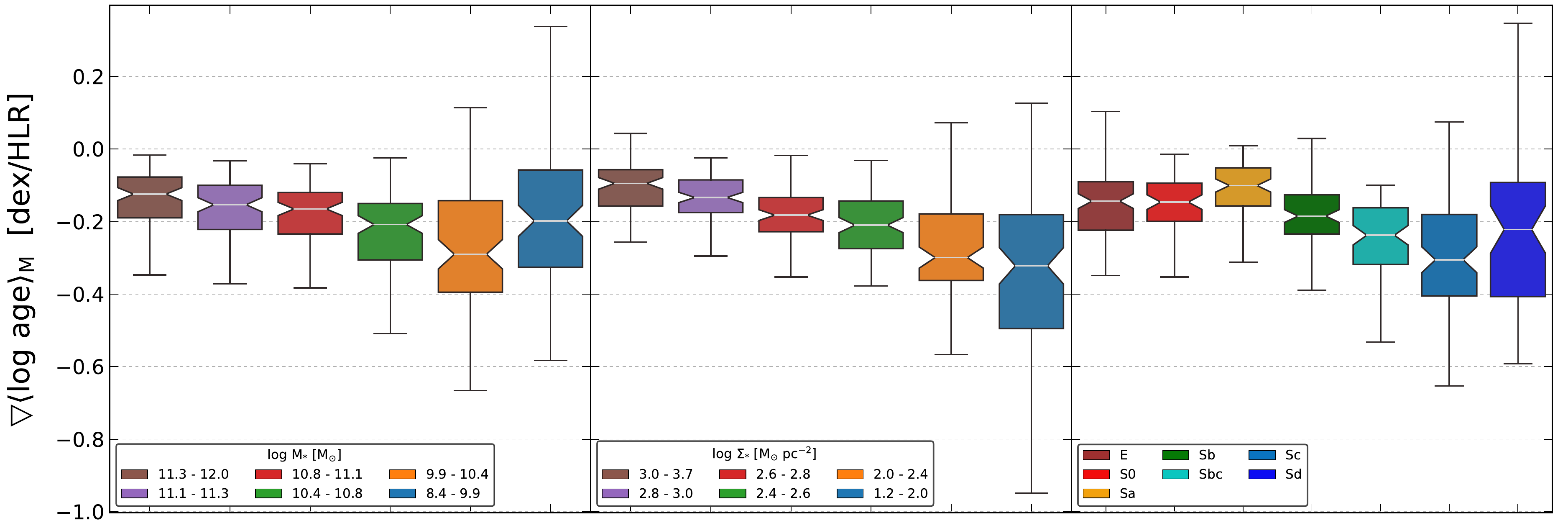}}
\caption{Box plot diagram of the mass-weighted age gradients \gagem\ (dex/\hlr) over the inner 1 \hlr\ for all 
galaxies as a function of stellar mass (left), \nmustar\ (center), and Hubble type (right). The gray horizontal 
line in each notched box plot shows the median of the distribution.}
\label{Fig:gat_mass__r}
\end{figure*}
%-----------------------------------------------------------------

%----------------------------------------------------------------- 
\begin{figure}
\centering
\resizebox{\hsize}{!}{\includegraphics{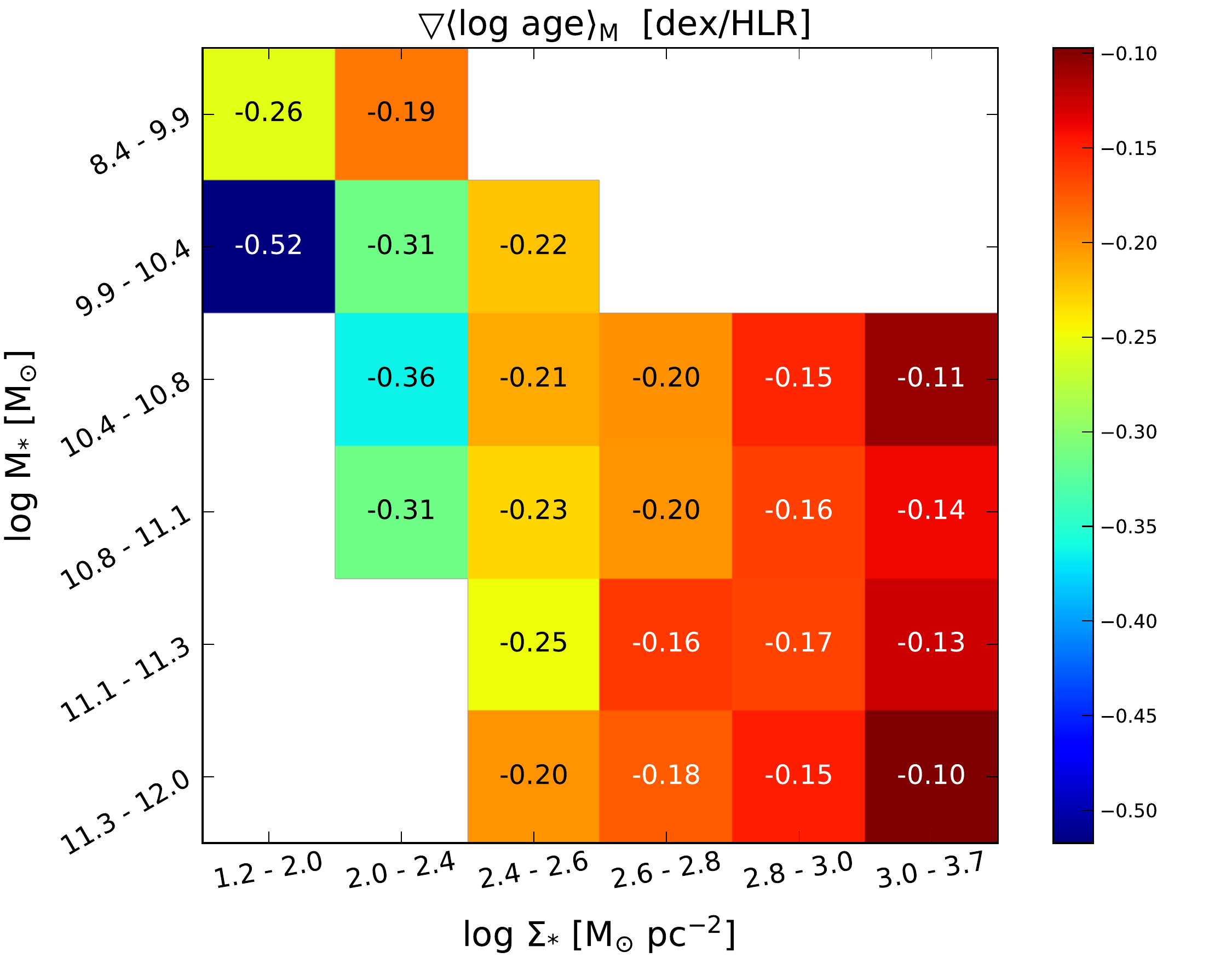}}
\caption{Visual representation of Table \ref{Tab:gagem} of the mass-weighted age gradients \gagem\ (dex/\hlr) 
estimated over the inner 1 \hlr\ for all galaxies as a function of {both} \mstar\ and \mustar.}
\label{Fig:gat_mcorsdmass__r}
\end{figure}
%-----------------------------------------------------------------

In \citet{GonzalezDelgado:2014a} we presented the results for the inner gradients in (light-weighted) age (and \mustar), 
and its relation with \mstar\ for 107 galaxies. The gradient of the light-weighted age was computed in the inner \hlr\ 
of each galaxy as the difference between the value of \agel\ at 1 \hlr\ and the value at the nucleus. In this work, 
we present the gradients of the {mass} weighted age \gagem\ as a function of both \mstar\ and \mustar\ in our 
defined bins. Here we follow a different approach for the calculation of the gradients.
We perform a robust linear fit over the entire inner 1 \hlr. The distribution of the slope of the fit for all galaxies 
as a function of stellar mass, \nmustar, and Hubble type is shown in Fig. \ref{Fig:gat_mass__r}.
On average, negative gradients are detected for all bins of the three parameters. A clear trend is seen both for \mstar\ 
(left panel) and \mustar\ (central panel) in the sense that the gradient is steeper (more negative) as the value 
of the parameter decreases. However, for the mass, the trend breaks in the lowest-mass bin where the average value 
increases with respect to the previous bin. The range of variation of the value of the gradients is particularly 
evident, as we have seen in previous plots, for the two lower bins, especially the lowest one, where a few galaxies 
show positive gradients, thus rising the average value of the distribution. As for the Hubble type (right panel), 
 two break points can be seen. The slope of early-type galaxies are very similar considering the dispersion, with 
average values of $-$0.15 dex/\hlr\ for E and $-$0.10 dex/\hlr\ for Sa galaxies, decreasing to the lowest value 
for Sc ($-$0.31 dex/\hlr), and then rising again for the Sd bin ($-$0.23 dex/\hlr), which presents the largest  
dispersion. Some of the Hubble-type bins also show a few positive galaxies as seen in the upper quartile of E, Sd, 
and particularly Sd galaxies. Again, the range of the interquartile for late-type galaxies (Sc and Sd) is wider than 
the other bins, which hints to a second dependence on other parameters, as seen in Sect. \ref{Sec:multiassembly}.
Most of the galaxies that show positive gradients have values compatible with flat or negative slopes  
taking into account their errors. The other very few cases that show steeper positive gradients are peculiar galaxies 
with a young star forming region in the center of the galaxy. On average, they do not show a particularly high sSFR 
value for their mass.

We explore now a possible multivariate dependence of the \gagem\ on both \mstar\ and \mustar. In this case, we perform 
a linear fit over the inner 1 \hlr\ of the averaged mass-weighted age profiles presented in Fig. \ref{Fig:at_mass__r}. 
The results are summarized in Table \ref{Tab:gagem} and visually represented in Fig. \ref{Fig:gat_mcorsdmass__r}. 
If we look at the values within a particular \mustar\ bin 
(columns), there is not a particular second dependence, within the errors, on the four largest \mstar\ 
bins. However, \mustar\ shows an important dependence on the low-mass range \mbin{8.4}{10.4}, where the slope 
becomes steeper (more negative) with increasing mass. This dependence disappears and the values of the slope fall 
into a plateau at the interface of the second and third \mstar\ bins ($\sim$ 10$^{10.4}$ \msun).
On the other hand, if we inspect the variation of \gagem\ with \mustar\ within a particular \mstar\ bin (rows), a 
clear trend is found. The slope flattens (less negative) with increasing values of \mustar. Thus, the wider range 
of values for the two lowest \mstar\ and \mustar\ bins seen in Fig. \ref{Fig:gat_mass__r} can be understood in 
terms of a stronger dependence on a second parameter in this regime.

As discussed in Sect. \ref{Sec:hmr_hlr}, stellar mass is more concentrated on average than light. We turn now to 
the question of whether or not the gradient results hold if the calculations are made based on the mass metric (\hmr) instead 
of a light metric (\hlr). Table \ref{Tab:gagem} shows the \gagem\ in the inner 1 \hmr\ for both \mstar\ and 
\mustar\ bins. The results are similar to those based on the light metric. On average, galaxies display negative 
gradients, but these less steep than the light counterpart. Similarly, the trend with parameters \mstar\ and \mustar\ 
also remains. These results indicate than the inner \hlr\ or \hmr\ gradients depend primarily on \mustar\, and 
to a lesser degree on \mstar\ for galaxies with \mustar\ $>$ 100 \mustarunits. As we have seen previously, the 
value of \mustar\ at 1 \hlr\ is an informative representative of the average \mustar\ for the whole galaxy. It is 
also a characteristic value for the \mustar\ of the disk. Low values of \mustar\ on the disk, because of the local 
\mustar\ - \agel\ relation, present lower values of \agel\ (a younger disk), and therefore, larger \agem\ gradients. 
These results can be understood in light of Fig. \ref{Fig:mgAp}: galaxies in the same \mstar\ bin 
display values of \tgin{80} but their \tgout{80} decrease with \mustar. This is the result of galaxies forming 
their disk (or envelope) at a later epoch.

%-----------------------------------------------------------------
\begin{table*}
\centering
\begin{tabular}{ccccccc}
\hline
\lmstar & \multicolumn{6}{c}{\lmustar} \\
            &         1.2 - 2.0 &         2.0 - 2.4 &         2.4 - 2.6 &         2.6 - 2.8 &         2.8 - 3.0 &         3.0 - 3.7 \\
\hline\hline
\multicolumn{7}{c}{\gagem\ [dex/\hlr]}\\        
\hline
% HLR
8.4 - 9.9                 &  -0.26 $\pm$ 0.04 &  -0.19 $\pm$ 0.02 &                -- &                -- &                -- &                -- \\
9.9 - 10.4                &  -0.52 $\pm$ 0.03 &  -0.31 $\pm$ 0.02 &  -0.22 $\pm$ 0.03 &                -- &                -- &                -- \\
10.4 - 10.8               &                -- &  -0.36 $\pm$ 0.03 &  -0.21 $\pm$ 0.02 &  -0.20 $\pm$ 0.02 &  -0.15 $\pm$ 0.02 &  -0.11 $\pm$ 0.03 \\
10.8 - 11.1               &                -- &  -0.31 $\pm$ 0.02 &  -0.23 $\pm$ 0.02 &  -0.20 $\pm$ 0.02 &  -0.16 $\pm$ 0.03 &  -0.14 $\pm$ 0.03 \\
11.1 - 11.3               &                -- &                -- &  -0.25 $\pm$ 0.03 &  -0.16 $\pm$ 0.02 &  -0.17 $\pm$ 0.03 &  -0.13 $\pm$ 0.03 \\
11.3 - 12.0               &                -- &                -- &  -0.20 $\pm$ 0.02 &  -0.18 $\pm$ 0.03 &  -0.15 $\pm$ 0.02 &  -0.10 $\pm$ 0.02 \\
\hline
\multicolumn{7}{c}{\gagem\ [dex/\hmr]}\\        
\hline
% HMR
8.4 - 9.9                 &  -0.19 $\pm$ 0.04 &  -0.15 $\pm$ 0.04 &                -- &                -- &                -- &                -- \\
9.9 - 10.4                &  -0.37 $\pm$ 0.03 &  -0.24 $\pm$ 0.03 &  -0.19 $\pm$ 0.02 &                -- &                -- &                -- \\
10.4 - 10.8               &                -- &  -0.22 $\pm$ 0.02 &  -0.17 $\pm$ 0.02 &  -0.14 $\pm$ 0.02 &  -0.11 $\pm$ 0.02 &  -0.10 $\pm$ 0.04 \\
10.8 - 11.1               &                -- &  -0.24 $\pm$ 0.02 &  -0.17 $\pm$ 0.02 &  -0.14 $\pm$ 0.02 &  -0.12 $\pm$ 0.03 &  -0.12 $\pm$ 0.02 \\
11.1 - 11.3               &                -- &                -- &  -0.17 $\pm$ 0.03 &  -0.11 $\pm$ 0.02 &  -0.15 $\pm$ 0.03 &  -0.09 $\pm$ 0.02 \\
11.3 - 12.0               &                -- &                -- &  -0.14 $\pm$ 0.03 &  -0.14 $\pm$ 0.03 &  -0.12 $\pm$ 0.02 &  -0.09 $\pm$ 0.02 \\
\hline
\end{tabular}
\centering
\caption{Mass-weighted age gradients \gagem\ of the inner 1 \hlr\ and inner 1 \hmr\ of the average radial profile of each stellar mass and \nmustar\ bin.}
\label{Tab:gagem}
\end{table*}
%-----------------------------------------------------------------

%%%%%%%%%%%%%%%%%%%%%%%%%%%%%%%%%%%%
%%%%%%%%%%%%% Section %%%%%%%%%%%%%%
%%%%%%%%%%%%%%%%%%%%%%%%%%%%%%%%%%%%
\section{Conclusions\label{Sec:conclusions}}

In this paper, we studied the mass assembly time scales of \ngal\ galaxies observed 
by CALIFA using the 3.5m telescope in Calar Alto. Our sample covers stellar masses from 
10$^{8.4}$ to 10$^{12}$ \msun\ (for a Salpeter IMF), including ellipticals, S0, and 
spirals from Sa to Sd. We applied the fossil record method of the stellar populations 
using the \stl\ code and a combination of SSP spectra from 
\citet{GonzalezDelgado:2005} plus \citet{Vazdekis:2015}. This base comprises eight 
metallicities from log log Z/Z$_{\odot}$ = -2.28 to +0.40, and 37 ages from 1 Myr 
to 14 Gyr. With the aid of our \pycasso\ pipeline we processed the spectral fitting 
results to produce 2D (R $\times$ t) maps of the mass growth and mass-weighted age, 
from which we obtain temporal and spatially resolved information. For each galaxy, 
the \agem\ maps are azimuthally averaged to produce radial profiles, both in units of 
\hlr\ and \hmr. All the analyzed quantities are stacked as a function of stellar mass, 
\nmustar, and Hubble type to explore the main trends.

Our main results are as follows:

\begin{enumerate}
\item Spatially resolved mass assembly. On average, galaxies form inside-out at any 
given stellar mass, \nmustar, and Hubble type bin, including the lowest-mass systems 
in our mass range. This result concerning low-mass galaxies could not be confirmed 
in our previous work \citet{Perez:2013}.
\item Galaxies show a significant diversity in their characteristic formation epochs 
for lower-mass systems. This diversity can be understood in terms of 
a strong dependence  of the mass assembly time scales on \mustar\ and Hubble type 
in the lower-mass range (\pow{8.4} to \pow{10.4}), while this dependence 
is very mild in higher-mass bins.
\item The lowest \hmr/\hlr\ ratio is found for galaxies between \pow{10.4} and \pow{11.1} 
\msun, where galaxies are 25\% smaller in mass than in light. Low-mass galaxies 
(\pow{8.4} - \pow{9.9} \msun) show the largest ratio with \hmr/\hlr $\sim$ 0.89. 
Sb and Sbc galaxies present the lowest \hmr/\hlr\ ratio (0.74). This results supports 
the result that galaxies grow inside-out.
\item At $z$ = 1, the downsizing pattern is, on average, spatially preserved. The inner 
parts of the most massive galaxies attained almost all their final mass ($\sim$ 97\%) 
and the envelopes were in the last stages of their assembly ($\sim$ 85\%). On the other 
hand, the inner parts of low-mass galaxies were at $\sim$ 70\%, while the outer 
parts close to half of their final value ($\sim$ 56\%).
\item We found, on average, negative \agem\ gradients in the inner 1 \hlr\ (or \hmr) for all 
galaxies. The radial profile flattens (slope less negative) with increasing values of \mustar. 
There is no significant dependence on \mstar\ within a particular \mustar\ bin, except for the 
lowest bin (\pow{8.4} - \pow{9.9} \msun), where the gradients become steeper.
\end{enumerate}

In summary, thanks to the large FoV of PPAK which covers galaxies in their entire optical extent, 
CALIFA has proven to be an excellent benchmark, in combination with the fossil record method, to 
test the formation and evolution of galaxies in terms of their spatially resolved properties. 
We have shown a complex multivariate dependence on stellar mass, \nmustar, and Hubble type of the 
mass assembly time scales and mass-weighted age gradients which once again demonstrates the 
convenience of these parameters in organizing the spatially resolved properties of galaxies.

\begin{acknowledgements}
CALIFA is the first legacy survey carried out at Calar Alto. The CALIFA collaboration would 
like to thank the IAA-CSIC and MPIA-MPG as major partners of the observatory, and CAHA itself, 
for the unique access to telescope time and support in manpower and infrastructures. We also 
thank the CAHA staff for their dedication to this project. We are grateful for the support of the IAA 
Computing group. 
Support from the Spanish Ministerio de Econom\'ia y Competitividad, through projects 
AYA2016-77846-P, AYA2014-57490-P, AYA2010-15081, and Junta de Andaluc\'ia P12-FQM-2828. 
SFS thanks PAPIIT-DGAPA-IA101217 project and CONACYT-IA-180125.
This research made use of Python 
(\href{http://www.python.org}{http://www.python.org}); Numpy \citep{vanDerWalt:2011}, 
Astropy \citep{Astropy:2013}, Pandas \citep{McKinney:2011}, Matplotlib \citep{Hunter:2007}, 
and Seaborn \citep{Waskom:2016}. We thank the referee for very useful comments that 
improved the presentation of the paper.
\end{acknowledgements}

% Bibliography
\bibliographystyle{aa}
\bibliography{rgb_califa_mgrowth}

\begin{thebibliography}{87}
\expandafter\ifx\csname natexlab\endcsname\relax\def\natexlab#1{#1}\fi

\bibitem[{{Abazajian} {et~al.}(2009){Abazajian}, {Adelman-McCarthy},
  {Ag{\"u}eros}, {Allam}, {Allende Prieto}, {An}, {Anderson}, {Anderson},
  {Annis}, {Bahcall}, \& et~al.}]{Abazajian:2009}
{Abazajian}, K.~N., {Adelman-McCarthy}, J.~K., {Ag{\"u}eros}, M.~A., {et~al.}
  2009, \apjs, 182, 543

\bibitem[{{Astropy Collaboration} {et~al.}(2013){Astropy Collaboration},
  {Robitaille}, {Tollerud}, {Greenfield}, {Droettboom}, {Bray}, {Aldcroft},
  {Davis}, {Ginsburg}, {Price-Whelan}, {Kerzendorf}, {Conley}, {Crighton},
  {Barbary}, {Muna}, {Ferguson}, {Grollier}, {Parikh}, {Nair}, {Unther},
  {Deil}, {Woillez}, {Conseil}, {Kramer}, {Turner}, {Singer}, {Fox}, {Weaver},
  {Zabalza}, {Edwards}, {Azalee Bostroem}, {Burke}, {Casey}, {Crawford},
  {Dencheva}, {Ely}, {Jenness}, {Labrie}, {Lian Lim}, {Pierfederici},
  {Pontzen}, {Ptak}, {Refsdal}, {Servillat}, \& {Streicher}}]{Astropy:2013}
{Astropy Collaboration}, {Robitaille}, T.~P., {Tollerud}, E.~J., {et~al.} 2013,
  \aap, 558, A33

\bibitem[{{Athanassoula} {et~al.}(2016){Athanassoula}, {Rodionov}, {Peschken},
  \& {Lambert}}]{Athanassoula:2016}
{Athanassoula}, E., {Rodionov}, S.~A., {Peschken}, N., \& {Lambert}, J.~C.
  2016, \apj, 821, 90

\bibitem[{{Aumer} \& {White}(2013)}]{Aumer:2013}
{Aumer}, M. \& {White}, S.~D.~M. 2013, \mnras, 428, 1055

\bibitem[{{Aumer} {et~al.}(2014){Aumer}, {White}, \& {Naab}}]{Aumer:2014}
{Aumer}, M., {White}, S.~D.~M., \& {Naab}, T. 2014, \mnras, 441, 3679

\bibitem[{{Avila-Reese} \& {Firmani}(2000)}]{AvilaReese:2000}
{Avila-Reese}, V. \& {Firmani}, C. 2000, \rmxaa, 36, 23

\bibitem[{{Baugh} {et~al.}(1996){Baugh}, {Cole}, \& {Frenk}}]{Baugh:1996}
{Baugh}, C.~M., {Cole}, S., \& {Frenk}, C.~S. 1996, \mnras, 283, 1361

\bibitem[{{Birnboim} {et~al.}(2007){Birnboim}, {Dekel}, \&
  {Neistein}}]{Birnboim:2007}
{Birnboim}, Y., {Dekel}, A., \& {Neistein}, E. 2007, \mnras, 380, 339

\bibitem[{{Brook} {et~al.}(2006){Brook}, {Kawata}, {Martel}, {Gibson}, \&
  {Bailin}}]{Brook:2006}
{Brook}, C.~B., {Kawata}, D., {Martel}, H., {Gibson}, B.~K., \& {Bailin}, J.
  2006, \apj, 639, 126

\bibitem[{{Brook} {et~al.}(2012){Brook}, {Stinson}, {Gibson}, {Kawata},
  {House}, {Miranda}, {Macci{\`o}}, {Pilkington}, {Ro{\v s}kar}, {Wadsley}, \&
  {Quinn}}]{Brook:2012}
{Brook}, C.~B., {Stinson}, G.~S., {Gibson}, B.~K., {et~al.} 2012, \mnras, 426,
  690

\bibitem[{{Buitrago} {et~al.}(2008){Buitrago}, {Trujillo}, {Conselice},
  {Bouwens}, {Dickinson}, \& {Yan}}]{Buitrago:2008}
{Buitrago}, F., {Trujillo}, I., {Conselice}, C.~J., {et~al.} 2008, \apjl, 687,
  L61

\bibitem[{{Bundy} {et~al.}(2015){Bundy}, {Bershady}, {Law}, {Yan}, {Drory},
  {MacDonald}, {Wake}, {Cherinka}, {S{\'a}nchez-Gallego}, {Weijmans}, {Thomas},
  {Tremonti}, {Masters}, {Coccato}, {Diamond-Stanic}, {Arag{\'o}n-Salamanca},
  {Avila-Reese}, {Badenes}, {Falc{\'o}n-Barroso}, {Belfiore}, {Bizyaev},
  {Blanc}, {Bland-Hawthorn}, {Blanton}, {Brownstein}, {Byler}, {Cappellari},
  {Conroy}, {Dutton}, {Emsellem}, {Etherington}, {Frinchaboy}, {Fu}, {Gunn},
  {Harding}, {Johnston}, {Kauffmann}, {Kinemuchi}, {Klaene}, {Knapen},
  {Leauthaud}, {Li}, {Lin}, {Maiolino}, {Malanushenko}, {Malanushenko}, {Mao},
  {Maraston}, {McDermid}, {Merrifield}, {Nichol}, {Oravetz}, {Pan}, {Parejko},
  {Sanchez}, {Schlegel}, {Simmons}, {Steele}, {Steinmetz}, {Thanjavur},
  {Thompson}, {Tinker}, {van den Bosch}, {Westfall}, {Wilkinson}, {Wright},
  {Xiao}, \& {Zhang}}]{Bundy:2015}
{Bundy}, K., {Bershady}, M.~A., {Law}, D.~R., {et~al.} 2015, \apj, 798, 7

\bibitem[{{Cappellari} \& {Copin}(2003)}]{Cappellari:2003}
{Cappellari}, M. \& {Copin}, Y. 2003, \mnras, 342, 345

\bibitem[{{Cardelli} {et~al.}(1989){Cardelli}, {Clayton}, \&
  {Mathis}}]{Cardelli:1989}
{Cardelli}, J.~A., {Clayton}, G.~C., \& {Mathis}, J.~S. 1989, \apj, 345, 245

\bibitem[{{Cid Fernandes} {et~al.}(2014){Cid Fernandes}, {Gonz{\'a}lez
  Delgado}, {Garc{\'{\i}}a Benito}, {P{\'e}rez}, {de Amorim}, {S{\'a}nchez},
  {Husemann}, {Falc{\'o}n Barroso}, {L{\'o}pez-Fern{\'a}ndez},
  {S{\'a}nchez-Bl{\'a}zquez}, {Vale Asari}, {Vazdekis}, {Walcher}, \&
  {Mast}}]{CidFernandes:2014}
{Cid Fernandes}, R., {Gonz{\'a}lez Delgado}, R.~M., {Garc{\'{\i}}a Benito}, R.,
  {et~al.} 2014, \aap, 561, A130

\bibitem[{{Cid Fernandes} {et~al.}(2005){Cid Fernandes}, {Mateus}, {Sodr{\'e}},
  {Stasi{\'n}ska}, \& {Gomes}}]{CidFernandes:2005}
{Cid Fernandes}, R., {Mateus}, A., {Sodr{\'e}}, L., {Stasi{\'n}ska}, G., \&
  {Gomes}, J.~M. 2005, \mnras, 358, 363

\bibitem[{{Cid Fernandes} {et~al.}(2013){Cid Fernandes}, {P{\'e}rez},
  {Garc{\'{\i}}a Benito}, {Gonz{\'a}lez Delgado}, {de Amorim}, {S{\'a}nchez},
  {Husemann}, {Falc{\'o}n Barroso}, {S{\'a}nchez-Bl{\'a}zquez}, {Walcher}, \&
  {Mast}}]{CidFernandes:2013}
{Cid Fernandes}, R., {P{\'e}rez}, E., {Garc{\'{\i}}a Benito}, R., {et~al.}
  2013, \aap, 557, A86

\bibitem[{{Croton} {et~al.}(2006){Croton}, {Springel}, {White}, {De Lucia},
  {Frenk}, {Gao}, {Jenkins}, {Kauffmann}, {Navarro}, \&
  {Yoshida}}]{Croton:2006}
{Croton}, D.~J., {Springel}, V., {White}, S.~D.~M., {et~al.} 2006, \mnras, 365,
  11

\bibitem[{{Dekel} {et~al.}(2009){Dekel}, {Birnboim}, {Engel}, {Freundlich},
  {Goerdt}, {Mumcuoglu}, {Neistein}, {Pichon}, {Teyssier}, \&
  {Zinger}}]{Dekel:2009}
{Dekel}, A., {Birnboim}, Y., {Engel}, G., {et~al.} 2009, \nat, 457, 451

\bibitem[{{Di Matteo} {et~al.}(2014){Di Matteo}, {Haywood}, {G{\'o}mez}, {van
  Damme}, {Combes}, {Hall{\'e}}, {Semelin}, {Lehnert}, \&
  {Katz}}]{diMatteo:2014}
{Di Matteo}, P., {Haywood}, M., {G{\'o}mez}, A., {et~al.} 2014, \aap, 567, A122

\bibitem[{{Dutton} {et~al.}(2010){Dutton}, {van den Bosch}, \&
  {Dekel}}]{Dutton:2010}
{Dutton}, A.~A., {van den Bosch}, F.~C., \& {Dekel}, A. 2010, \mnras, 405, 1690

\bibitem[{{Elmegreen} {et~al.}(2008){Elmegreen}, {Bournaud}, \&
  {Elmegreen}}]{Elmegreen:2008}
{Elmegreen}, B.~G., {Bournaud}, F., \& {Elmegreen}, D.~M. 2008, \apj, 688, 67

\bibitem[{{Elmegreen} {et~al.}(2007){Elmegreen}, {Elmegreen}, {Ravindranath},
  \& {Coe}}]{Elmegreen:2007}
{Elmegreen}, D.~M., {Elmegreen}, B.~G., {Ravindranath}, S., \& {Coe}, D.~A.
  2007, \apj, 658, 763

\bibitem[{{Fisher} \& {Drory}(2010)}]{Fisher:2010}
{Fisher}, D.~B. \& {Drory}, N. 2010, \apj, 716, 942

\bibitem[{{Fontanot} {et~al.}(2009){Fontanot}, {De Lucia}, {Monaco},
  {Somerville}, \& {Santini}}]{Fontanot:2009}
{Fontanot}, F., {De Lucia}, G., {Monaco}, P., {Somerville}, R.~S., \&
  {Santini}, P. 2009, \mnras, 397, 1776

\bibitem[{{F{\"o}rster Schreiber} {et~al.}(2011){F{\"o}rster Schreiber},
  {Shapley}, {Erb}, {Genzel}, {Steidel}, {Bouch{\'e}}, {Cresci}, \&
  {Davies}}]{ForsterScheiber:2011}
{F{\"o}rster Schreiber}, N.~M., {Shapley}, A.~E., {Erb}, D.~K., {et~al.} 2011,
  \apj, 731, 65

\bibitem[{{Garc{\'{\i}}a-Benito} {et~al.}(2015){Garc{\'{\i}}a-Benito},
  {Zibetti}, {S{\'a}nchez}, {Husemann}, {de Amorim}, {Castillo-Morales}, {Cid
  Fernandes}, {Ellis}, {Falc{\'o}n-Barroso}, {Galbany}, {Gil de Paz},
  {Gonz{\'a}lez Delgado}, {Lacerda}, {L{\'o}pez-Fernandez}, {de
  Lorenzo-C{\'a}ceres}, {Lyubenova}, {Marino}, {Mast}, {Mendoza}, {P{\'e}rez},
  {Vale Asari}, {Aguerri}, {Ascasibar}, {Bekerait*error*{\.e}},
  {Bland-Hawthorn}, {Barrera-Ballesteros}, {Bomans}, {Cano-D{\'{\i}}az},
  {Catal{\'a}n-Torrecilla}, {Cortijo}, {Delgado-Inglada}, {Demleitner},
  {Dettmar}, {D{\'{\i}}az}, {Florido}, {Gallazzi}, {Garc{\'{\i}}a-Lorenzo},
  {Gomes}, {Holmes}, {Iglesias-P{\'a}ramo}, {Jahnke}, {Kalinova}, {Kehrig},
  {Kennicutt}, {L{\'o}pez-S{\'a}nchez}, {M{\'a}rquez}, {Masegosa}, {Meidt},
  {Mendez-Abreu}, {Moll{\'a}}, {Monreal-Ibero}, {Morisset}, {del Olmo},
  {Papaderos}, {P{\'e}rez}, {Quirrenbach}, {Rosales-Ortega}, {Roth},
  {Ruiz-Lara}, {S{\'a}nchez-Bl{\'a}zquez}, {S{\'a}nchez-Menguiano}, {Singh},
  {Spekkens}, {Stanishev}, {Torres-Papaqui}, {van de Ven}, {Vilchez},
  {Walcher}, {Wild}, {Wisotzki}, {Ziegler}, {Alves}, {Barrado}, {Quintana}, \&
  {Aceituno}}]{dr2}
{Garc{\'{\i}}a-Benito}, R., {Zibetti}, S., {S{\'a}nchez}, S.~F., {et~al.} 2015,
  \aap, 576, A135

\bibitem[{{Genzel} {et~al.}(2008){Genzel}, {Burkert}, {Bouch{\'e}}, {Cresci},
  {F{\"o}rster Schreiber}, {Shapley}, {Shapiro}, {Tacconi}, {Buschkamp},
  {Cimatti}, {Daddi}, {Davies}, {Eisenhauer}, {Erb}, {Genel}, {Gerhard},
  {Hicks}, {Lutz}, {Naab}, {Ott}, {Rabien}, {Renzini}, {Steidel}, {Sternberg},
  \& {Lilly}}]{Genzel:2008}
{Genzel}, R., {Burkert}, A., {Bouch{\'e}}, N., {et~al.} 2008, \apj, 687, 59

\bibitem[{{Genzel} {et~al.}(2014){Genzel}, {F{\"o}rster Schreiber}, {Lang},
  {Tacchella}, {Tacconi}, {Wuyts}, {Bandara}, {Burkert}, {Buschkamp},
  {Carollo}, {Cresci}, {Davies}, {Eisenhauer}, {Hicks}, {Kurk}, {Lilly},
  {Lutz}, {Mancini}, {Naab}, {Newman}, {Peng}, {Renzini}, {Shapiro Griffin},
  {Sternberg}, {Vergani}, {Wisnioski}, {Wuyts}, \& {Zamorani}}]{Genzel:2014}
{Genzel}, R., {F{\"o}rster Schreiber}, N.~M., {Lang}, P., {et~al.} 2014, \apj,
  785, 75

\bibitem[{{Gil de Paz} {et~al.}(2007){Gil de Paz}, {Boissier}, {Madore},
  {Seibert}, {Joe}, {Boselli}, {Wyder}, {Thilker}, {Bianchi}, {Rey}, {Rich},
  {Barlow}, {Conrow}, {Forster}, {Friedman}, {Martin}, {Morrissey}, {Neff},
  {Schiminovich}, {Small}, {Donas}, {Heckman}, {Lee}, {Milliard}, {Szalay}, \&
  {Yi}}]{GildePaz:2007}
{Gil de Paz}, A., {Boissier}, S., {Madore}, B.~F., {et~al.} 2007, \apjs, 173,
  185

\bibitem[{{Gil de Paz} {et~al.}(2005){Gil de Paz}, {Madore}, {Boissier},
  {Swaters}, {Popescu}, {Tuffs}, {Sheth}, {Kennicutt}, {Bianchi}, {Thilker}, \&
  {Martin}}]{GildePaz:2005}
{Gil de Paz}, A., {Madore}, B.~F., {Boissier}, S., {et~al.} 2005, \apjl, 627,
  L29

\bibitem[{{Goddard} {et~al.}(2016){Goddard}, {Thomas}, {Maraston}, {Westfall},
  {Etherington}, {Riffel}, {Mallmann}, {Zheng}, {Argudo-Fernandez}, {Lian},
  {Bershady}, {Bundy}, {Drory}, {Law}, {Yan}, {Wake}, {Weijmans}, {Bizyaev},
  {Brownstein}, {Lane}, {Maiolino}, {Masters}, {Merrifield}, {Nitschelm},
  {Pan}, {Roman-Lopes}, {Storchi-Bergmann}, \& {Schneider}}]{Goddard:2016}
{Goddard}, D., {Thomas}, D., {Maraston}, C., {et~al.} 2016, ArXiv e-prints
  [\eprint[arXiv]{1612.01546}]

\bibitem[{{Gonz{\'a}lez Delgado} {et~al.}(2005){Gonz{\'a}lez Delgado},
  {Cervi{\~n}o}, {Martins}, {Leitherer}, \&
  {Hauschildt}}]{GonzalezDelgado:2005}
{Gonz{\'a}lez Delgado}, R.~M., {Cervi{\~n}o}, M., {Martins}, L.~P.,
  {Leitherer}, C., \& {Hauschildt}, P.~H. 2005, \mnras, 357, 945

\bibitem[{{Gonz{\'a}lez Delgado} {et~al.}(2014{\natexlab{a}}){Gonz{\'a}lez
  Delgado}, {Cid Fernandes}, {Garc{\'{\i}}a-Benito}, {P{\'e}rez}, {de Amorim},
  {Cortijo-Ferrero}, {Lacerda}, {L{\'o}pez Fern{\'a}ndez}, {S{\'a}nchez}, {Vale
  Asari}, {Alves}, {Bland-Hawthorn}, {Galbany}, {Gallazzi}, {Husemann},
  {Bekeraite}, {Jungwiert}, {L{\'o}pez-S{\'a}nchez}, {de Lorenzo-C{\'a}ceres},
  {Marino}, {Mast}, {Moll{\'a}}, {del Olmo}, {S{\'a}nchez-Bl{\'a}zquez}, {van
  de Ven}, {V{\'{\i}}lchez}, {Walcher}, {Wisotzki}, {Ziegler}, \&
  {Collaboration920}}]{GonzalezDelgado:2014b}
{Gonz{\'a}lez Delgado}, R.~M., {Cid Fernandes}, R., {Garc{\'{\i}}a-Benito}, R.,
  {et~al.} 2014{\natexlab{a}}, \apjl, 791, L16

\bibitem[{{Gonz{\'a}lez Delgado} {et~al.}(2016){Gonz{\'a}lez Delgado}, {Cid
  Fernandes}, {P{\'e}rez}, {Garc{\'{\i}}a-Benito}, {L{\'o}pez Fern{\'a}ndez},
  {Lacerda}, {Cortijo-Ferrero}, {de Amorim}, {Vale Asari}, {S{\'a}nchez},
  {Walcher}, {Wisotzki}, {Mast}, {Alves}, {Ascasibar}, {Bland-Hawthorn},
  {Galbany}, {Kennicutt}, {M{\'a}rquez}, {Masegosa}, {Moll{\'a}},
  {S{\'a}nchez-Bl{\'a}zquez}, \& {V{\'{\i}}lchez}}]{GonzalezDelgado:2016}
{Gonz{\'a}lez Delgado}, R.~M., {Cid Fernandes}, R., {P{\'e}rez}, E., {et~al.}
  2016, \aap, 590, A44

\bibitem[{{Gonz{\'a}lez Delgado} {et~al.}(2015){Gonz{\'a}lez Delgado},
  {Garc{\'{\i}}a-Benito}, {P{\'e}rez}, {Cid Fernandes}, {de Amorim},
  {Cortijo-Ferrero}, {Lacerda}, {L{\'o}pez Fern{\'a}ndez}, {Vale-Asari},
  {S{\'a}nchez}, {Moll{\'a}}, {Ruiz-Lara}, {S{\'a}nchez-Bl{\'a}zquez},
  {Walcher}, {Alves}, {Aguerri}, {Bekerait{\'e}}, {Bland-Hawthorn}, {Galbany},
  {Gallazzi}, {Husemann}, {Iglesias-P{\'a}ramo}, {Kalinova},
  {L{\'o}pez-S{\'a}nchez}, {Marino}, {M{\'a}rquez}, {Masegosa}, {Mast},
  {M{\'e}ndez-Abreu}, {Mendoza}, {del Olmo}, {P{\'e}rez}, {Quirrenbach}, \&
  {Zibetti}}]{GonzalezDelgado:2015}
{Gonz{\'a}lez Delgado}, R.~M., {Garc{\'{\i}}a-Benito}, R., {P{\'e}rez}, E.,
  {et~al.} 2015, \aap, 581, A103

\bibitem[{{Gonz{\'a}lez Delgado} {et~al.}(2014{\natexlab{b}}){Gonz{\'a}lez
  Delgado}, {P{\'e}rez}, {Cid Fernandes}, {Garc{\'{\i}}a-Benito}, {de Amorim},
  {S{\'a}nchez}, {Husemann}, {Cortijo-Ferrero}, {L{\'o}pez Fern{\'a}ndez},
  {S{\'a}nchez-Bl{\'a}zquez}, {Bekeraite}, {Walcher}, {Falc{\'o}n-Barroso},
  {Gallazzi}, {van de Ven}, {Alves}, {Bland-Hawthorn}, {Kennicutt}, {Kupko},
  {Lyubenova}, {Mast}, {Moll{\'a}}, {Marino}, {Quirrenbach}, {V{\'{\i}}lchez},
  \& {Wisotzki}}]{GonzalezDelgado:2014a}
{Gonz{\'a}lez Delgado}, R.~M., {P{\'e}rez}, E., {Cid Fernandes}, R., {et~al.}
  2014{\natexlab{b}}, \aap, 562, A47

\bibitem[{{Gonz{\'a}lez Delgado et al.}(2017)}]{GonzalezDelgado:2017}
{Gonz{\'a}lez Delgado et al.}, R.~M. 2017, \aap, submitted

\bibitem[{{Haywood} {et~al.}(2013){Haywood}, {Di Matteo}, {Lehnert}, {Katz}, \&
  {G{\'o}mez}}]{Haywood:2013}
{Haywood}, M., {Di Matteo}, P., {Lehnert}, M.~D., {Katz}, D., \& {G{\'o}mez},
  A. 2013, \aap, 560, A109

\bibitem[{{Haywood} {et~al.}(2015){Haywood}, {Di Matteo}, {Snaith}, \&
  {Lehnert}}]{Haywood:2015}
{Haywood}, M., {Di Matteo}, P., {Snaith}, O., \& {Lehnert}, M.~D. 2015, \aap,
  579, A5

\bibitem[{{Hilz} {et~al.}(2013){Hilz}, {Naab}, \& {Ostriker}}]{Hilz:2013}
{Hilz}, M., {Naab}, T., \& {Ostriker}, J.~P. 2013, \mnras, 429, 2924

\bibitem[{{Hopkins} {et~al.}(2009){Hopkins}, {Cox}, {Younger}, \&
  {Hernquist}}]{Hopkins:2009}
{Hopkins}, P.~F., {Cox}, T.~J., {Younger}, J.~D., \& {Hernquist}, L. 2009,
  \apj, 691, 1168

\bibitem[{Hunter(2007)}]{Hunter:2007}
Hunter, J.~D. 2007, Computing In Science \& Engineering, 9, 90

\bibitem[{{Ibarra-Medel} {et~al.}(2016){Ibarra-Medel}, {S{\'a}nchez},
  {Avila-Reese}, {Hern{\'a}ndez-Toledo}, {Gonz{\'a}lez}, {Drory}, {Bundy},
  {Bizyaev}, {Cano-D{\'{\i}}az}, {Malanushenko}, {Pan}, {Roman-Lopes}, \&
  {Thomas}}]{IbarraMedel:2016}
{Ibarra-Medel}, H.~J., {S{\'a}nchez}, S.~F., {Avila-Reese}, V., {et~al.} 2016,
  \mnras, 463, 2799

\bibitem[{{Kelz} {et~al.}(2006){Kelz}, {Verheijen}, {Roth}, {Bauer}, {Becker},
  {Paschke}, {Popow}, {S{\'a}nchez}, \& {Laux}}]{Kelz:2006}
{Kelz}, A., {Verheijen}, M.~A.~W., {Roth}, M.~M., {et~al.} 2006, \pasp, 118,
  129

\bibitem[{{Larson}(1976)}]{Larson:1976}
{Larson}, R.~B. 1976, \mnras, 176, 31

\bibitem[{{Lilly} {et~al.}(2013){Lilly}, {Carollo}, {Pipino}, {Renzini}, \&
  {Peng}}]{Lilly:2013}
{Lilly}, S.~J., {Carollo}, C.~M., {Pipino}, A., {Renzini}, A., \& {Peng}, Y.
  2013, \apj, 772, 119

\bibitem[{{Marino} {et~al.}(2016){Marino}, {Gil de Paz}, {S{\'a}nchez},
  {S{\'a}nchez-Bl{\'a}zquez}, {Cardiel}, {Castillo-Morales}, {Pascual},
  {V{\'{\i}}lchez}, {Kehrig}, {Moll{\'a}}, {Mendez-Abreu},
  {Catal{\'a}n-Torrecilla}, {Florido}, {Perez}, {Ruiz-Lara}, {Ellis},
  {L{\'o}pez-S{\'a}nchez}, {Gonz{\'a}lez Delgado}, {de Lorenzo-C{\'a}ceres},
  {Garc{\'{\i}}a-Benito}, {Galbany}, {Zibetti}, {Cortijo}, {Kalinova}, {Mast},
  {Iglesias-P{\'a}ramo}, {Papaderos}, {Walcher}, \&
  {Bland-Hawthorn}}]{Marino:2016}
{Marino}, R.~A., {Gil de Paz}, A., {S{\'a}nchez}, S.~F., {et~al.} 2016, \aap,
  585, A47

\bibitem[{{Martig} {et~al.}(2009){Martig}, {Bournaud}, {Teyssier}, \&
  {Dekel}}]{Martig:2009}
{Martig}, M., {Bournaud}, F., {Teyssier}, R., \& {Dekel}, A. 2009, \apj, 707,
  250

\bibitem[{{Martin} {et~al.}(2005){Martin}, {Fanson}, {Schiminovich},
  {Morrissey}, {Friedman}, {Barlow}, {Conrow}, {Grange}, {Jelinsky},
  {Milliard}, {Siegmund}, {Bianchi}, {Byun}, {Donas}, {Forster}, {Heckman},
  {Lee}, {Madore}, {Malina}, {Neff}, {Rich}, {Small}, {Surber}, {Szalay},
  {Welsh}, \& {Wyder}}]{Martin:2005}
{Martin}, D.~C., {Fanson}, J., {Schiminovich}, D., {et~al.} 2005, \apjl, 619,
  L1

\bibitem[{McKinney(2011)}]{McKinney:2011}
McKinney, W. 2011, pandas : powerful Python data analysis toolkit

\bibitem[{{Mo} {et~al.}(1998){Mo}, {Mao}, \& {White}}]{Mo:1998}
{Mo}, H.~J., {Mao}, S., \& {White}, S.~D.~M. 1998, \mnras, 295, 319

\bibitem[{{Mu{\~n}oz-Mateos} {et~al.}(2007){Mu{\~n}oz-Mateos}, {Gil de Paz},
  {Boissier}, {Zamorano}, {Jarrett}, {Gallego}, \& {Madore}}]{MunozMateos:2007}
{Mu{\~n}oz-Mateos}, J.~C., {Gil de Paz}, A., {Boissier}, S., {et~al.} 2007,
  \apj, 658, 1006

\bibitem[{{Naab} {et~al.}(2009){Naab}, {Johansson}, \& {Ostriker}}]{Naab:2009}
{Naab}, T., {Johansson}, P.~H., \& {Ostriker}, J.~P. 2009, \apjl, 699, L178

\bibitem[{{Neistein} {et~al.}(2006){Neistein}, {van den Bosch}, \&
  {Dekel}}]{Neistein:2006}
{Neistein}, E., {van den Bosch}, F.~C., \& {Dekel}, A. 2006, \mnras, 372, 933

\bibitem[{{Nelson} {et~al.}(2012){Nelson}, {van Dokkum}, {Brammer},
  {F{\"o}rster Schreiber}, {Franx}, {Fumagalli}, {Patel}, {Rix}, {Skelton},
  {Bezanson}, {Da Cunha}, {Kriek}, {Labbe}, {Lundgren}, {Quadri}, \&
  {Schmidt}}]{Nelson:2012}
{Nelson}, E.~J., {van Dokkum}, P.~G., {Brammer}, G., {et~al.} 2012, \apjl, 747,
  L28

\bibitem[{{Nelson} {et~al.}(2016){Nelson}, {van Dokkum}, {F{\"o}rster
  Schreiber}, {Franx}, {Brammer}, {Momcheva}, {Wuyts}, {Whitaker}, {Skelton},
  {Fumagalli}, {Hayward}, {Kriek}, {Labb{\'e}}, {Leja}, {Rix}, {Tacconi}, {van
  der Wel}, {van den Bosch}, {Oesch}, {Dickey}, \& {Ulf Lange}}]{Nelson:2016a}
{Nelson}, E.~J., {van Dokkum}, P.~G., {F{\"o}rster Schreiber}, N.~M., {et~al.}
  2016, \apj, 828, 27

\bibitem[{{Oser} {et~al.}(2010){Oser}, {Ostriker}, {Naab}, {Johansson}, \&
  {Burkert}}]{Oser:2010}
{Oser}, L., {Ostriker}, J.~P., {Naab}, T., {Johansson}, P.~H., \& {Burkert}, A.
  2010, \apj, 725, 2312

\bibitem[{{Patel} {et~al.}(2013){Patel}, {van Dokkum}, {Franx}, {Quadri},
  {Muzzin}, {Marchesini}, {Williams}, {Holden}, \& {Stefanon}}]{Patel:2013}
{Patel}, S.~G., {van Dokkum}, P.~G., {Franx}, M., {et~al.} 2013, \apj, 766, 15

\bibitem[{{P{\'e}rez} {et~al.}(2013){P{\'e}rez}, {Cid Fernandes}, {Gonz{\'a}lez
  Delgado}, {Garc{\'{\i}}a-Benito}, {S{\'a}nchez}, {Husemann}, {Mast},
  {Rod{\'o}n}, {Kupko}, {Backsmann}, {de Amorim}, {van de Ven}, {Walcher},
  {Wisotzki}, {Cortijo-Ferrero}, \& {collaboration6}}]{Perez:2013}
{P{\'e}rez}, E., {Cid Fernandes}, R., {Gonz{\'a}lez Delgado}, R.~M., {et~al.}
  2013, \apjl, 764, L1

\bibitem[{{Pezzulli} {et~al.}(2015){Pezzulli}, {Fraternali}, {Boissier}, \&
  {Mu{\~n}oz-Mateos}}]{Pezzulli:2015}
{Pezzulli}, G., {Fraternali}, F., {Boissier}, S., \& {Mu{\~n}oz-Mateos}, J.~C.
  2015, \mnras, 451, 2324

\bibitem[{{Pilkington} {et~al.}(2012){Pilkington}, {Few}, {Gibson}, {Calura},
  {Michel-Dansac}, {Thacker}, {Moll{\'a}}, {Matteucci}, {Rahimi}, {Kawata},
  {Kobayashi}, {Brook}, {Stinson}, {Couchman}, {Bailin}, \&
  {Wadsley}}]{Pilkington:2012a}
{Pilkington}, K., {Few}, C.~G., {Gibson}, B.~K., {et~al.} 2012, \aap, 540, A56

\bibitem[{{Roth} {et~al.}(2005){Roth}, {Kelz}, {Fechner}, {Hahn}, {Bauer},
  {Becker}, {B{\"o}hm}, {Christensen}, {Dionies}, {Paschke}, {Popow}, {Wolter},
  {Schmoll}, {Laux}, \& {Altmann}}]{Roth:2005}
{Roth}, M.~M., {Kelz}, A., {Fechner}, T., {et~al.} 2005, \pasp, 117, 620

\bibitem[{{S{\'a}nchez} {et~al.}(2016){S{\'a}nchez}, {Garc{\'{\i}}a-Benito},
  {Zibetti}, {Walcher}, {Husemann}, {Mendoza}, {Galbany}, {Falc{\'o}n-Barroso},
  {Mast}, {Aceituno}, {Aguerri}, {Alves}, {Amorim}, {Ascasibar},
  {Barrado-Navascues}, {Barrera-Ballesteros}, {Bekerait{\`e}},
  {Bland-Hawthorn}, {Cano D{\'{\i}}az}, {Cid Fernandes}, {Cavichia}, {Cortijo},
  {Dannerbauer}, {Demleitner}, {D{\'{\i}}az}, {Dettmar}, {de
  Lorenzo-C{\'a}ceres}, {del Olmo}, {Galazzi}, {Garc{\'{\i}}a-Lorenzo}, {Gil de
  Paz}, {Gonz{\'a}lez Delgado}, {Holmes}, {Igl{\'e}sias-P{\'a}ramo}, {Kehrig},
  {Kelz}, {Kennicutt}, {Kleemann}, {Lacerda}, {L{\'o}pez Fern{\'a}ndez},
  {L{\'o}pez S{\'a}nchez}, {Lyubenova}, {Marino}, {M{\'a}rquez},
  {Mendez-Abreu}, {Moll{\'a}}, {Monreal-Ibero}, {Ortega Minakata},
  {Torres-Papaqui}, {P{\'e}rez}, {Rosales-Ortega}, {Roth},
  {S{\'a}nchez-Bl{\'a}zquez}, {Schilling}, {Spekkens}, {Vale Asari}, {van den
  Bosch}, {van de Ven}, {Vilchez}, {Wild}, {Wisotzki}, {Y{\i}ld{\i}r{\i}m}, \&
  {Ziegler}}]{dr3}
{S{\'a}nchez}, S.~F., {Garc{\'{\i}}a-Benito}, R., {Zibetti}, S., {et~al.} 2016,
  \aap, 594, A36

\bibitem[{{S{\'a}nchez} {et~al.}(2012{\natexlab{a}}){S{\'a}nchez}, {Kennicutt},
  {Gil de Paz}, {van de Ven}, {V{\'{\i}}lchez}, {Wisotzki}, {Walcher}, {Mast},
  {Aguerri}, {Albiol-P{\'e}rez}, {Alonso-Herrero}, {Alves}, {Bakos},
  {Bart{\'a}kov{\'a}}, {Bland-Hawthorn}, {Boselli}, {Bomans},
  {Castillo-Morales}, {Cortijo-Ferrero}, {de Lorenzo-C{\'a}ceres}, {Del Olmo},
  {Dettmar}, {D{\'{\i}}az}, {Ellis}, {Falc{\'o}n-Barroso}, {Flores},
  {Gallazzi}, {Garc{\'{\i}}a-Lorenzo}, {Gonz{\'a}lez Delgado}, {Gruel},
  {Haines}, {Hao}, {Husemann}, {Igl{\'e}sias-P{\'a}ramo}, {Jahnke}, {Johnson},
  {Jungwiert}, {Kalinova}, {Kehrig}, {Kupko}, {L{\'o}pez-S{\'a}nchez},
  {Lyubenova}, {Marino}, {M{\'a}rmol-Queralt{\'o}}, {M{\'a}rquez}, {Masegosa},
  {Meidt}, {Mendez-Abreu}, {Monreal-Ibero}, {Montijo}, {Mour{\~a}o},
  {Palacios-Navarro}, {Papaderos}, {Pasquali}, {Peletier}, {P{\'e}rez},
  {P{\'e}rez}, {Quirrenbach}, {Rela{\~n}o}, {Rosales-Ortega}, {Roth},
  {Ruiz-Lara}, {S{\'a}nchez-Bl{\'a}zquez}, {Sengupta}, {Singh}, {Stanishev},
  {Trager}, {Vazdekis}, {Viironen}, {Wild}, {Zibetti}, \& {Ziegler}}]{califa}
{S{\'a}nchez}, S.~F., {Kennicutt}, R.~C., {Gil de Paz}, A., {et~al.}
  2012{\natexlab{a}}, \aap, 538, A8

\bibitem[{{S{\'a}nchez} {et~al.}(2014){S{\'a}nchez}, {Rosales-Ortega},
  {Iglesias-P{\'a}ramo}, {Moll{\'a}}, {Barrera-Ballesteros}, {Marino},
  {P{\'e}rez}, {S{\'a}nchez-Blazquez}, {Gonz{\'a}lez Delgado}, {Cid Fernandes},
  {de Lorenzo-C{\'a}ceres}, {Mendez-Abreu}, {Galbany}, {Falcon-Barroso},
  {Miralles-Caballero}, {Husemann}, {Garc{\'{\i}}a-Benito}, {Mast}, {Walcher},
  {Gil de Paz}, {Garc{\'{\i}}a-Lorenzo}, {Jungwiert}, {V{\'{\i}}lchez},
  {J{\'{\i}}lkov{\'a}}, {Lyubenova}, {Cortijo-Ferrero}, {D{\'{\i}}az},
  {Wisotzki}, {M{\'a}rquez}, {Bland-Hawthorn}, {Ellis}, {van de Ven}, {Jahnke},
  {Papaderos}, {Gomes}, {Mendoza}, \& {L{\'o}pez-S{\'a}nchez}}]{Sanchez:2014}
{S{\'a}nchez}, S.~F., {Rosales-Ortega}, F.~F., {Iglesias-P{\'a}ramo}, J.,
  {et~al.} 2014, \aap, 563, A49

\bibitem[{{S{\'a}nchez} {et~al.}(2012{\natexlab{b}}){S{\'a}nchez},
  {Rosales-Ortega}, {Marino}, {Iglesias-P{\'a}ramo}, {V{\'{\i}}lchez},
  {Kennicutt}, {D{\'{\i}}az}, {Mast}, {Monreal-Ibero}, {Garc{\'{\i}}a-Benito},
  {Bland-Hawthorn}, {P{\'e}rez}, {Gonz{\'a}lez Delgado}, {Husemann},
  {L{\'o}pez-S{\'a}nchez}, {Cid Fernandes}, {Kehrig}, {Walcher}, {Gil de Paz},
  \& {Ellis}}]{Sanchez:2012}
{S{\'a}nchez}, S.~F., {Rosales-Ortega}, F.~F., {Marino}, R.~A., {et~al.}
  2012{\natexlab{b}}, \aap, 546, A2

\bibitem[{{S{\'a}nchez-Bl{\'a}zquez} {et~al.}(2014){S{\'a}nchez-Bl{\'a}zquez},
  {Rosales-Ortega}, {M{\'e}ndez-Abreu}, {P{\'e}rez}, {S{\'a}nchez}, {Zibetti},
  {Aguerri}, {Bland-Hawthorn}, {Catal{\'a}n-Torrecilla}, {Cid Fernandes}, {de
  Amorim}, {de Lorenzo-Caceres}, {Falc{\'o}n-Barroso}, {Galazzi},
  {Garc{\'{\i}}a Benito}, {Gil de Paz}, {Gonz{\'a}lez Delgado}, {Husemann},
  {Iglesias-P{\'a}ramo}, {Jungwiert}, {Marino}, {M{\'a}rquez}, {Mast},
  {Mendoza}, {Moll{\'a}}, {Papaderos}, {Ruiz-Lara}, {van de Ven}, {Walcher}, \&
  {Wisotzki}}]{SanchezBlazquez:2014}
{S{\'a}nchez-Bl{\'a}zquez}, P., {Rosales-Ortega}, F.~F., {M{\'e}ndez-Abreu},
  J., {et~al.} 2014, \aap, 570, A6

\bibitem[{{S{\'a}nchez-Menguiano} {et~al.}(2016){S{\'a}nchez-Menguiano},
  {S{\'a}nchez}, {P{\'e}rez}, {Garc{\'{\i}}a-Benito}, {Husemann}, {Mast},
  {Mendoza}, {Ruiz-Lara}, {Ascasibar}, {Bland-Hawthorn}, {Cavichia},
  {D{\'{\i}}az}, {Florido}, {Galbany}, {G{\'o}nzalez Delgado}, {Kehrig},
  {Marino}, {M{\'a}rquez}, {Masegosa}, {M{\'e}ndez-Abreu}, {Moll{\'a}}, {Del
  Olmo}, {P{\'e}rez}, {S{\'a}nchez-Bl{\'a}zquez}, {Stanishev}, {Walcher},
  {L{\'o}pez-S{\'a}nchez}, \& {Califa Collaboration}}]{SanchezMenguiano:2016}
{S{\'a}nchez-Menguiano}, L., {S{\'a}nchez}, S.~F., {P{\'e}rez}, I., {et~al.}
  2016, \aap, 587, A70

\bibitem[{{Springel} \& {Hernquist}(2005)}]{Springel:2005}
{Springel}, V. \& {Hernquist}, L. 2005, \apjl, 622, L9

\bibitem[{{Steinmetz} \& {Navarro}(2002)}]{Steinmetz:2002}
{Steinmetz}, M. \& {Navarro}, J.~F. 2002, \na, 7, 155

\bibitem[{{Szomoru} {et~al.}(2012){Szomoru}, {Franx}, \& {van
  Dokkum}}]{Szomoru:2012}
{Szomoru}, D., {Franx}, M., \& {van Dokkum}, P.~G. 2012, \apj, 749, 121

\bibitem[{{Szomoru} {et~al.}(2010){Szomoru}, {Franx}, {van Dokkum}, {Trenti},
  {Illingworth}, {Labb{\'e}}, {Bouwens}, {Oesch}, \& {Carollo}}]{Szomoru:2010}
{Szomoru}, D., {Franx}, M., {van Dokkum}, P.~G., {et~al.} 2010, \apjl, 714,
  L244

\bibitem[{{Szomoru} {et~al.}(2013){Szomoru}, {Franx}, {van Dokkum}, {Trenti},
  {Illingworth}, {Labb{\'e}}, \& {Oesch}}]{Szomoru:2013}
{Szomoru}, D., {Franx}, M., {van Dokkum}, P.~G., {et~al.} 2013, \apj, 763, 73

\bibitem[{{Tinsley}(1968)}]{Tinsley:1968}
{Tinsley}, B.~M. 1968, \apj, 151, 547

\bibitem[{{Toomre} \& {Toomre}(1972)}]{Toomre:1972}
{Toomre}, A. \& {Toomre}, J. 1972, \apj, 178, 623

\bibitem[{{Trujillo} {et~al.}(2006){Trujillo}, {F{\"o}rster Schreiber},
  {Rudnick}, {Barden}, {Franx}, {Rix}, {Caldwell}, {McIntosh}, {Toft},
  {H{\"a}ussler}, {Zirm}, {van Dokkum}, {Labb{\'e}}, {Moorwood},
  {R{\"o}ttgering}, {van der Wel}, {van der Werf}, \& {van
  Starkenburg}}]{Trujillo:2006}
{Trujillo}, I., {F{\"o}rster Schreiber}, N.~M., {Rudnick}, G., {et~al.} 2006,
  \apj, 650, 18

\bibitem[{Van Der~Walt {et~al.}(2011)Van Der~Walt, Colbert, \&
  Varoquaux}]{vanDerWalt:2011}
Van Der~Walt, S., Colbert, S.~C., \& Varoquaux, G. 2011, Computing in Science
  \& Engineering, 13, 22

\bibitem[{{van Dokkum} {et~al.}(2013){van Dokkum}, {Leja}, {Nelson}, {Patel},
  {Skelton}, {Momcheva}, {Brammer}, {Whitaker}, {Lundgren}, {Fumagalli},
  {Conroy}, {F{\"o}rster Schreiber}, {Franx}, {Kriek}, {Labb{\'e}},
  {Marchesini}, {Rix}, {van der Wel}, \& {Wuyts}}]{vanDokkum:2013}
{van Dokkum}, P.~G., {Leja}, J., {Nelson}, E.~J., {et~al.} 2013, \apjl, 771,
  L35

\bibitem[{{van Dokkum} {et~al.}(2010){van Dokkum}, {Whitaker}, {Brammer},
  {Franx}, {Kriek}, {Labb{\'e}}, {Marchesini}, {Quadri}, {Bezanson},
  {Illingworth}, {Muzzin}, {Rudnick}, {Tal}, \& {Wake}}]{vanDokkum:2010}
{van Dokkum}, P.~G., {Whitaker}, K.~E., {Brammer}, G., {et~al.} 2010, \apj,
  709, 1018

\bibitem[{{Vazdekis} {et~al.}(2015){Vazdekis}, {Coelho}, {Cassisi},
  {Ricciardelli}, {Falc{\'o}n-Barroso}, {S{\'a}nchez-Bl{\'a}zquez}, {Barbera},
  {Beasley}, \& {Pietrinferni}}]{Vazdekis:2015}
{Vazdekis}, A., {Coelho}, P., {Cassisi}, S., {et~al.} 2015, \mnras, 449, 1177

\bibitem[{{Verheijen} {et~al.}(2004){Verheijen}, {Bershady}, {Andersen},
  {Swaters}, {Westfall}, {Kelz}, \& {Roth}}]{Verheijen:2004}
{Verheijen}, M.~A.~W., {Bershady}, M.~A., {Andersen}, D.~R., {et~al.} 2004,
  Astronomische Nachrichten, 325, 151

\bibitem[{{Walcher} {et~al.}(2014){Walcher}, {Wisotzki}, {Bekerait{\'e}},
  {Husemann}, {Iglesias-P{\'a}ramo}, {Backsmann}, {Barrera Ballesteros},
  {Catal{\'a}n-Torrecilla}, {Cortijo}, {del Olmo}, {Garcia Lorenzo},
  {Falc{\'o}n-Barroso}, {Jilkova}, {Kalinova}, {Mast}, {Marino},
  {M{\'e}ndez-Abreu}, {Pasquali}, {S{\'a}nchez}, {Trager}, {Zibetti},
  {Aguerri}, {Alves}, {Bland-Hawthorn}, {Boselli}, {Castillo Morales}, {Cid
  Fernandes}, {Flores}, {Galbany}, {Gallazzi}, {Garc{\'{\i}}a-Benito}, {Gil de
  Paz}, {Gonz{\'a}lez-Delgado}, {Jahnke}, {Jungwiert}, {Kehrig}, {Lyubenova},
  {M{\'a}rquez Perez}, {Masegosa}, {Monreal Ibero}, {P{\'e}rez}, {Quirrenbach},
  {Rosales-Ortega}, {Roth}, {Sanchez-Blazquez}, {Spekkens}, {Tundo}, {van de
  Ven}, {Verheijen}, {Vilchez}, \& {Ziegler}}]{Walcher:2014}
{Walcher}, C.~J., {Wisotzki}, L., {Bekerait{\'e}}, S., {et~al.} 2014, \aap,
  569, A1 (W14)

\bibitem[{Waskom {et~al.}(2016)Waskom, Botvinnik, drewokane, Hobson, Halchenko,
  Lukauskas, Warmenhoven, Cole, Hoyer, Vanderplas, gkunter, Villalba, Quintero,
  Martin, Miles, Meyer, Augspurger, Yarkoni, Bachant, Evans, Fitzgerald, Nagy,
  Ziegler, Megies, Wehner, St-Jean, Coelho, Hitz, Lee, \& Rocher}]{Waskom:2016}
Waskom, M., Botvinnik, O., drewokane, {et~al.} 2016, seaborn: v0.7.0 (January
  2016)

\bibitem[{{Wilkinson} {et~al.}(2015){Wilkinson}, {Maraston}, {Thomas},
  {Coccato}, {Tojeiro}, {Cappellari}, {Belfiore}, {Bershady}, {Blanton},
  {Bundy}, {Cales}, {Cherinka}, {Drory}, {Emsellem}, {Fu}, {Law}, {Li},
  {Maiolino}, {Masters}, {Tremonti}, {Wake}, {Wang}, {Weijmans}, {Xiao}, {Yan},
  {Zhang}, {Bizyaev}, {Brinkmann}, {Kinemuchi}, {Malanushenko}, {Malanushenko},
  {Oravetz}, {Pan}, \& {Simmons}}]{Wilkinson:2015}
{Wilkinson}, D.~M., {Maraston}, C., {Thomas}, D., {et~al.} 2015, \mnras, 449,
  328

\bibitem[{{Wuyts} {et~al.}(2013){Wuyts}, {F{\"o}rster Schreiber}, {Nelson},
  {van Dokkum}, {Brammer}, {Chang}, {Faber}, {Ferguson}, {Franx}, {Fumagalli},
  {Genzel}, {Grogin}, {Kocevski}, {Koekemoer}, {Lundgren}, {Lutz}, {McGrath},
  {Momcheva}, {Rosario}, {Skelton}, {Tacconi}, {van der Wel}, \&
  {Whitaker}}]{Wuyts:2013}
{Wuyts}, S., {F{\"o}rster Schreiber}, N.~M., {Nelson}, E.~J., {et~al.} 2013,
  \apj, 779, 135

\bibitem[{{Zheng} {et~al.}(2017){Zheng}, {Wang}, {Ge}, {Mao}, {Li}, {Li}, {Mo},
  {Goddard}, {Bundy}, {Li}, {Nair}, {Lin}, {Long}, {Riffel}, {Thomas},
  {Masters}, {Bizyaev}, {Brownstein}, {Zhang}, {Law}, {Drory}, {Roman Lopes},
  \& {Malanushenko}}]{Zheng:2017}
{Zheng}, Z., {Wang}, H., {Ge}, J., {et~al.} 2017, \mnras, 465, 4572

\end{thebibliography}

\end{document}